\pdfoutput=1
\RequirePackage{ifpdf}
\ifpdf 
\documentclass[pdftex]{sigma}
\else
\documentclass{sigma}
\fi

\usepackage{multirow}

\DeclareMathOperator{\sn}{sn}

\begin{document}

\allowdisplaybreaks

\renewcommand{\PaperNumber}{014}

\FirstPageHeading

\ShortArticleName{Integrable Boundary for Quad-Graph Systems}

\ArticleName{Integrable Boundary for Quad-Graph Systems:\\
Three-Dimensional Boundary Consistency}

\Author{Vincent CAUDRELIER~$^\dag$, Nicolas CRAMP\'E~$^\ddag$ and Qi Cheng ZHANG~$^\dag$}

\AuthorNameForHeading{V.~Caudrelier, N.~Cramp\'e and Q.C.~Zhang}

\Address{$^\dag$~Department of Mathematical Science, City University London,\\
\hphantom{$^\dag$}~Northampton Square, London EC1V 0HB, UK}
\EmailD{\href{mailto:v.caudrelier@city.ac.uk}{v.caudrelier@city.ac.uk},
\href{mailto:c.zhang@leeds.ac.uk}{c.zhang@leeds.ac.uk}}

\Address{$^\ddag$~CNRS, Laboratoire Charles Coulomb, UMR 5221,\\
\hphantom{$^\ddag$}~Place Eug\`ene Bataillon -- CC070, F-34095 Montpellier, France}
\EmailD{\href{mailto:nicolas.crampe@um2.fr}{nicolas.crampe@um2.fr}}

\ArticleDates{Received July 19, 2013, in f\/inal form February 05, 2014; Published online February 12, 2014}

\Abstract{We propose the notion of integrable boundary in the context of discrete integrable systems on
quad-graphs.
The equation characterizing the boundary must satisfy a~compatibility equation with the one characterizing
the bulk that we called the three-dimensional~(3D) boundary consistency.
In comparison to the usual~3D consistency condition which is linked to a~cube, our 3D boundary
consistency condition lives on a~half of a~rhombic dodecahedron.
The We provide a~list of integrable boundaries associated to each quad-graph equation of the
classif\/ication obtained by Adler, Bobenko and Suris.
Then, the use of the term ``integrable boundary''
is justif\/ied by the facts that there are B\"acklund
transformations and a~zero curvature representation for systems with boundary satisfying our condition.
We discuss the three-leg form of boundary equations, obtain associated discrete Toda-type models with
boundary and recover previous results as particular cases.
Finally, the connection between the 3D boundary consistency and the set-theoretical ref\/lection equation
is established.}

\Keywords{discrete integrable systems; quad-graph equations; 3D-consistency; B\"acklund transformations;
zero curvature representation; Toda-type systems; set-theoretical ref\/lection equation}

\Classification{05C10; 37K10; 39A12; 57M15}

\section{Introduction}

Discrete integrable systems arise from various motivations in applied or pure mathematics like the need to
preserve integrability of certains continuous equations when performing numerical (and hence discrete)
simulations or the theory of discrete dif\/ferential geometry.
In previous work concerning an important class of such systems known as integrable quad-graph
equations~\cite{BS2}, one of the original motivations was to study discrete dif\/ferential geometry of
surfaces.
In this context, a~general construction allows one to always obtain a~discretization of the surface in
terms of quadrilaterals~\cite{BS2}, at least as long as one is only concerned with the \textit{bulk} of the
surface and does not worry about its boundary (if it has one).
The vertices of the graph thus obtained can be seen as discrete space-time points where the f\/ield is
attached.
The dynamics of the f\/ield is then specif\/ied by an equation of motion involving the values of the
f\/ield at the four vertices forming an elementary quadrilateral.
Typically, this is of the form
\begin{gather}
\label{eq_Q}
Q(u_{00},u_{10},u_{01},u_{11};a,b)=0,
\end{gather}
where $u_{00}$, $u_{10}$, $u_{01}$, $u_{11}$ are the values of the f\/ield at the vertices of the quadrilateral and~$a$,~$b$ are parameters (see l.h.s.\ of the Fig.~\ref{quad_tri}).

There exist several integrability criteria that dif\/ferent authors use to characterize the notion of
discrete integrability.
Let us mention for instance algebraic entropy~\cite{VB}, singularity conf\/inement~\cite{GRP} or the 3D
consistency/consistency around the cube condition~\cite{BS2,Nij}.
The latter is deeply related to the notion of discrete Lax pair, discrete B\"acklund transformations and
other classical notions of integrability and, combined with a~few other assumptions, led to the important
ABS classif\/ication of quad-graph equations~\cite{ABS}.
This fact and the similarity of the above structures with those existing in the case of continuous
integrable systems make it a~popular criterion.

In this paper, we want to propose a~way of def\/ining an integrable discrete system on a~quad-graph with
\textit{boundary} as arising from the discretization of a~surface while taking into account its boundary.
From the geometric point of view, this is a~natural generalization of the discrete dif\/ferential geometry
motivation to the case of a~surface with boundary.
From the point of view of discrete integrable quad-graph equations, this then allows us to tackle the
problem of formulating the analog of the 3D consistency condition together with its consequences, i.e.\
B\"acklund transformations and zero curvature formulation.
We also introduce Toda-type systems with boundary through the three-leg form of integrable equations on
quad-graphs and we recover the previous approach to boundary conditions for discrete integrable systems
presented in~\cite{HK}.

We give a~precise def\/inition of the discretization procedure and of a~quad-graph with boundary in
Section~\ref{sec:qgb}.
Then we show how to def\/ine a~discrete system on it.
In addition to the elementary quadrilateral and the corresponding equation of motion~\eqref{eq_Q}, the new
crucial ingredient is an elementary triangle together with the corresponding boundary equation of the form
\begin{gather*}
q(x,y,z;a)=0,
\end{gather*}
where $x$, $y$, $z$ are the values of the f\/ield at the vertices of the triangle, and a~function $\sigma$
determining the ef\/fect of the boundary on the labelling (see r.h.s.\ of the Fig.~\ref{quad_tri}).
We provide our def\/inition of integrability in this context by def\/ining the \textit{3D boundary
consistency condition} involving~$Q$,~$q$ and $\sigma$ in Section~\ref{sec:integrability} and then
go on to present a~method that allows us to f\/ind solutions for~$q$ and~$\sigma$ for a~given~$Q$ in
Section~\ref{sec:class}.
In Section~\ref{sec:aspects}, we justify further our introduction of the 3D boundary consistency
condition by discussing B\"acklund transformations and the zero curvature representation in the presence of
a~boundary.
Section~\ref{sec:toda} introduces the three-leg form for boundary equations and we use this to
def\/ine systems of Toda-type with boundary.
As an example, we recover as particular case the approach to boundary conditions of~\cite{HK} in (fully)
discrete integrable systems.
Finally, in Section~\ref{sec:conn} we establish the connection between the 3D boundary consistency
condition introduced in this paper and the set-theoretical ref\/lection equation~\cite{CCZ, VZ2} in the same
spirit~\cite{PTV} as the 3D consistency condition is related to the set-theoretical Yang--Baxter
equation~\cite{Drin}.
Conclusions and outlooks are collected in the last section.

\section{Integrable quad-graph systems with boundary}

In this section, we f\/irst def\/ine the notion of quad-graph with boundary and then use it to def\/ine the
elementary blocks needed to study integrable equations on quad-graphs with boundary.

\subsection{Quad-graph with boundary
\label{sec:qgb}
}

Our starting point is the def\/inition of a~quad-graph from a~cellular decomposition of an oriented surface
${\cal S}$ containing only quadrilateral faces.
As explained in~\cite{BS2}, a~quad-graph can always be obtained from an arbitrary cellular
decomposition~${\cal G}$ by forming the double~${\cal D}$ (in the sense of~\cite{BS2}, otherwise called
a~diamond in~\cite{M}) of~${\cal G}$ and its dual cellular decomposition~${\cal G}^*$.
So far, these notions apply to surfaces without a~boundary.
For the case of a~surface with boundary, the notion of dual cellular decomposition does not exist in
general.
In~\cite{M}, Def\/inition~1 gives the def\/inition of an object~$\Gamma^*$ associated to a~cellular
decomposition~$\Gamma$ of a~compact surface with boundary and it is noted that~$\Gamma^*$ is not a~cellular
decomposition of the surface.
Nevertheless, in Section~C.3 of~\cite{mercatthesis},
a~notion of double is given for a~surface with
boundary (and not necessarily compact, like a~half-plane) and we adapt it here for our purposes.
In particular, we will see that the generic structure that arises is what we call a~quad-graph with
boundary with its faces being either quadrilaterals or ``half quadrilaterals'', i.e.\ triangles.

\begin{figure}[t]\centering
\includegraphics{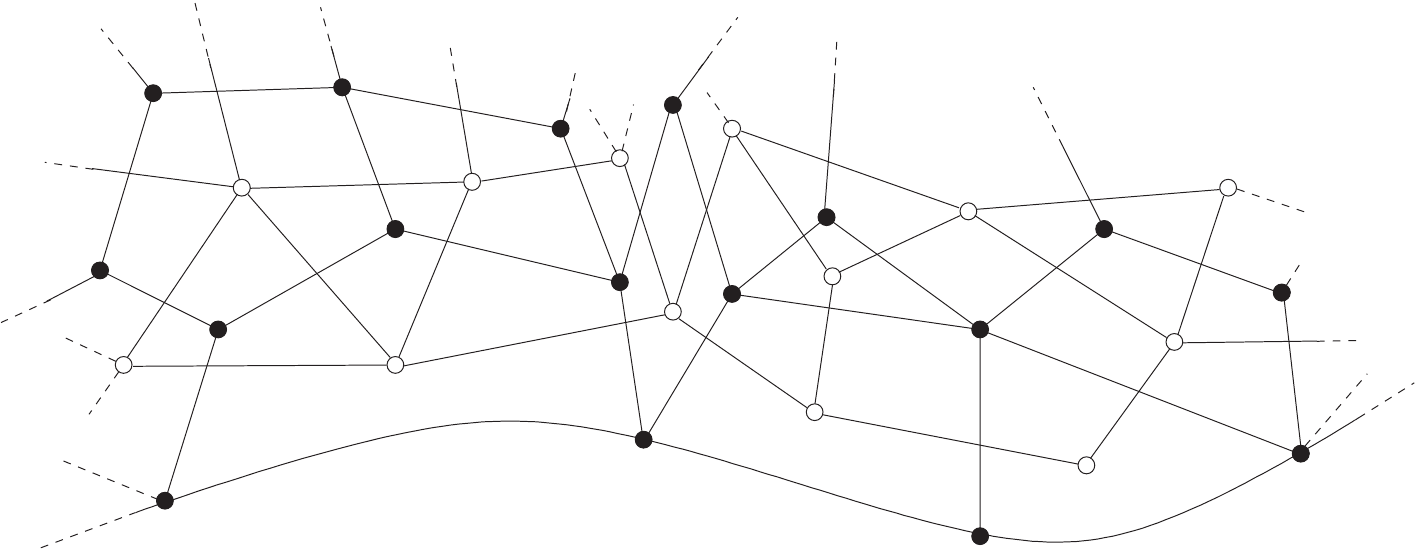}
\caption{Example of a~cellular decomposition ${\cal
G}$ and of the associated~${\cal G}^*$.
The ``horizontal'' curved line is the boundary of the underlying surface~${\cal S}$.
The black dots and the straight lines connecting them are the vertices and edges of the initial cellular
decomposition~${\cal G}$.
The white dots and the straight lines connecting them are the vertices and edges of~${\cal G}^*$.
Edges of~${\cal G}$ on the boundary and the boundary itself are identif\/ied in this picture.}
\label{graphdual}
\end{figure}

So let us consider a~cellular decomposition ${\cal G}$ of our surface ${\cal S}$ with a~boundary.
We denote, respectively, by $F$, $E$ and $V$ the set of faces, edges and vertices of this cellular
decomposition.
Following~\cite{mercatthesis} and adapting slightly, we def\/ine ${\cal G}^*$ as the following collection
of cells with~$F^*$,~$E^*$ and~$V^*$ respectively the sets of faces, edges and vertices.
\begin{itemize}
\itemsep=0pt
\item To each face in $F$, we associate a~vertex~$v^*$ in~$V^*$ (called white), placed inside
the face.
\item To each edge $e\in E$ not on the boundary, we associate the dual edge $e^*\in E^*$ which cuts it
tranversally and forms a~path between the two white vertices contained in the two adjacents faces in $F$
separated by~$e$.
\item To each vertex $v\in V$ (called black) not on the boundary, we associate the face in~$F^*$ that
contains it, i.e.\ the face whose boundary is made of the edges in~$E^*$ that cross the edges in $E$ having
the vertex $v$ under consideration as one of their ends.
\end{itemize}
Compared to Def\/inition~6 of~\cite{mercatthesis}, in our def\/inition of ${\cal G}^*$, we include neither
the dual edge corresponding to an edge on the boundary nor the additional white vertex in the middle of an
edge in $E$ belonging to the boundary.
A typical conf\/iguration of~${\cal G}$ and ${\cal G}^*$ is shown in Fig.~\ref{graphdual}.

\begin{figure}[t]\centering
\includegraphics{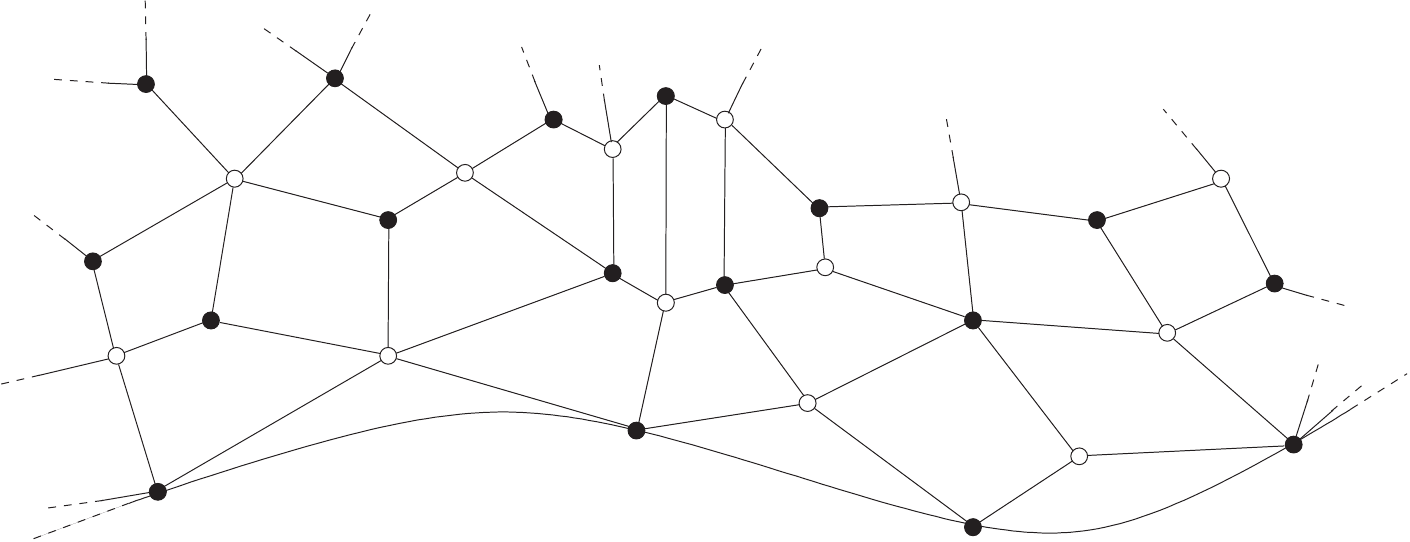}
\caption{Example of a~quad-graph with boundary~${\cal D}$.
Black and white dots are the vertices of the double and the straight lines are edges.
Edges of~${\cal G}$ on the boundary and the boundary itself are identif\/ied in this picture and form the
rest of the edges of~${\cal D}$.
Faces are therefore either quadrilaterals (with two black and two white alternating vertices) or triangles
lying along the boundary (with two black vertices and one white vertex).}
\label{fig:diamond}
\end{figure}

We now form the structure that we will call a~quad-graph with boundary below.
Let us denote it ${\cal D}$.
The vertices of ${\cal D}$ are all the black and white vertices, i.e.\ $V_{\cal D}=V\cup V^*$.
The edges of ${\cal D}$ are all the edges in $E$ that lie on the boundary of ${\cal S}$ together with those
edges $(w,b)$ obtained by connecting a~white vertex to each of the black vertices sitting on the face that
contains the white vertex\footnote{Note that these edges are neither in $E$ nor $E^*$.}.
The faces are then taken to be the interiors of the ``polygons'' thus obtained.
The result of this procedure for ${\cal G}$ and ${\cal G}^*$ as in Fig.~\ref{graphdual} is shown in
Fig.~\ref{fig:diamond}.
The faces are therefore of two types:
\begin{itemize}\itemsep=0pt

\item Quadrilaterals, with two black and two white vertices, in the bulk.

\item Triangles, with two black vertices (on the boundary) and one white vertex (inside ${\cal S}$),
alongside the boundary.

\end{itemize}

\begin{definition}
A quad-graph with boundary is the collection of vertices and edges of ${\cal D}$ obtained as described
above from a~cellular decomposition of a~surface with boundary.
\end{definition}

As an important by-product, this procedure to get a~quad-graph with boundary allows one to obtain
a~bipartite graph\footnote{To be precise, we should not include the boundary edges if we want to talk about
a~bipartite graph.\label{footnote_bi}}.
In addition, all the vertices on the boundary are of the same type.
This property will be important in the construction of the Toda models in Section~\ref{sec:toda}.

\subsection{Discrete equations on quad-graph with boundary}

\begin{figure}[t]\centering
\includegraphics{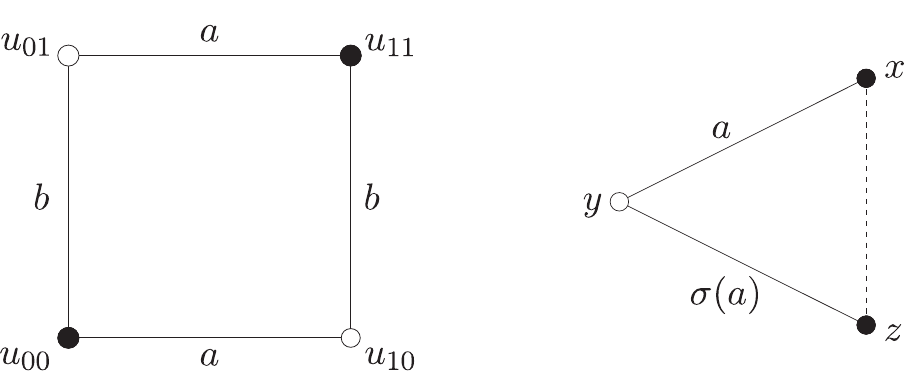}
\caption{Elementary blocks to construct a~discrete system with boundary.}\label{quad_tri}
\end{figure}

We are now ready to def\/ine discrete equations on a~quad-graph with boundary.
As usual, we associate a~f\/ield to this quad-graph (i.e.\ a function from $V\cup V^*$, the set of
vertices, to~${\mathbb C}$) and a~constraint between the values of the f\/ield around a~face.
This constraint can be seen as the equation of motion for the f\/ield.
For each quadrilateral face, the constraint is, as usual, def\/ined~by
\begin{gather}
\label{eq:Q}
Q(u_{00},u_{10},u_{01},u_{11};a,b)=0,
\end{gather}
where $u_{00}, u_{10}, u_{01}, u_{11}\in {\mathbb C}$ are the values of the f\/ield at each vertex around the
face and $a,b\in{\mathbb C}$ are parameters associated to opposite edges.
One usually represents this equation as on the l.h.s.\ of Fig.~\ref{quad_tri}.
We will refer to equation~\eqref{eq:Q} as describing the bulk dynamics.
Sometimes, one demands additional properties~\cite{BS2} for the function
$Q(u_{00},u_{10},u_{01},u_{11};a,b)$ such as linearity on each variable $u_{00}$, $u_{10}$,  $u_{01}$, $u_{11}$
(af\/f\/ine-linearity), symmetry by exchange of these variables ($D_4$-symmetry), the tetrahedron property
or the existence of the three-leg form (see Section~\ref{sec:toda}).

\begin{figure}[t]\centering
\includegraphics{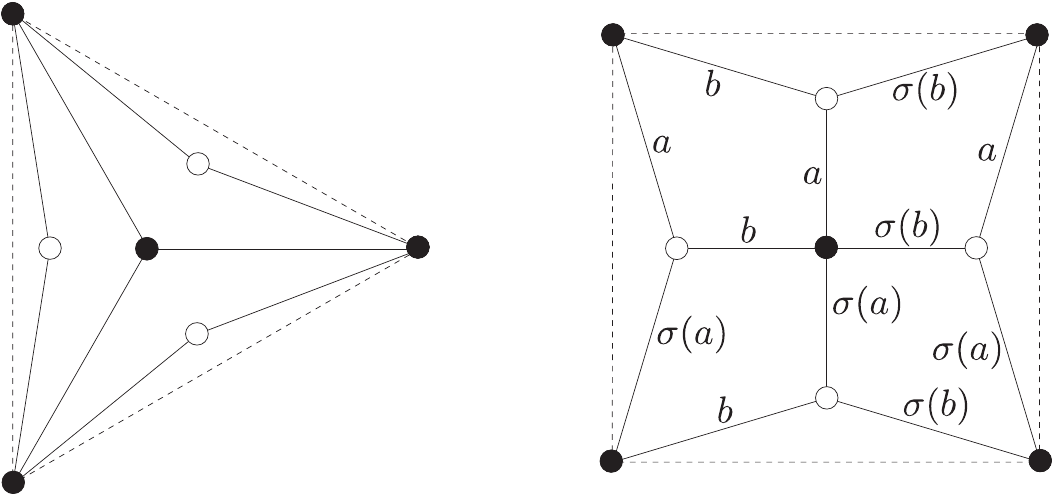}
\caption{Two examples of quad-graphs with boundary
(represented by a~dashed line here).
Only the one of the right-hand side can be labelled according to our rules for quadrilaterals and
triangles.}\label{fig:trc}
\end{figure}

The new elementary building block needed to def\/ine a~discrete system on a~quad-graph with boundary is an
equation of the following type def\/ined on each triangular face
\begin{gather}
\label{eq:q}
q(x,y,z;a)=0,
\end{gather}
where $x$, $z$ are values of the f\/ield on the boundary, $y$ a~value inside the surface and~$a$ is
a~parameter associated to one edge (the other edge is associated to a~function $\sigma(a)$ of the parameter~$a$).
We represent this equation as on the r.h.s.\ of Fig.~\ref{quad_tri} where the dashed line represents the
edge on the boundary of the surface.
We will refer to this equation as describing the boundary dynamics.
In general, there is no special requirement on $q$ or $\sigma$ but we will see that our methods to
construct expressions for $q$ result in $q$ having certain properties:
\begin{itemize}
\itemsep=0pt \item Linearity: the function $q(x,y,z;a)$ is linear in the variables $x$ and $z$.
Let us emphasize that it may not be linear in $y$.
\item Symmetry: one has $\sigma(\sigma(a))=a$ and
\begin{gather*}
q(x,y,z;a)=0
\quad
\Leftrightarrow
\quad
q(z,y,x;\sigma(a))=0.
\end{gather*}
\item In some cases, a~three-leg form for $q$ inherited from the three-leg form of the corresponding bulk
equation $Q=0$.
We go back to this point in Section~\ref{sec:toda}.
\end{itemize}
Let us remark that for a~given quad-graph with boundary (and for $\sigma$ being not the identity), it is
not always possible to label the edges with parameters as prescribed above.
We show on the l.h.s.\ of Fig.~\ref{fig:trc} a~quad-graph with boundary for which we cannot f\/ind
a~suitable labelling.
On the r.h.s., we show a~case where this is possible.
In the following, we consider only those quad-graph with boundary that can be labelled.
It would be very interesting to study this combinatorial problem in general but this goes beyond the scope
of this paper.

\subsection{Integrability: the 3D boundary consistency condition}\label{sec:integrability}

\begin{figure}[t]\centering
\includegraphics{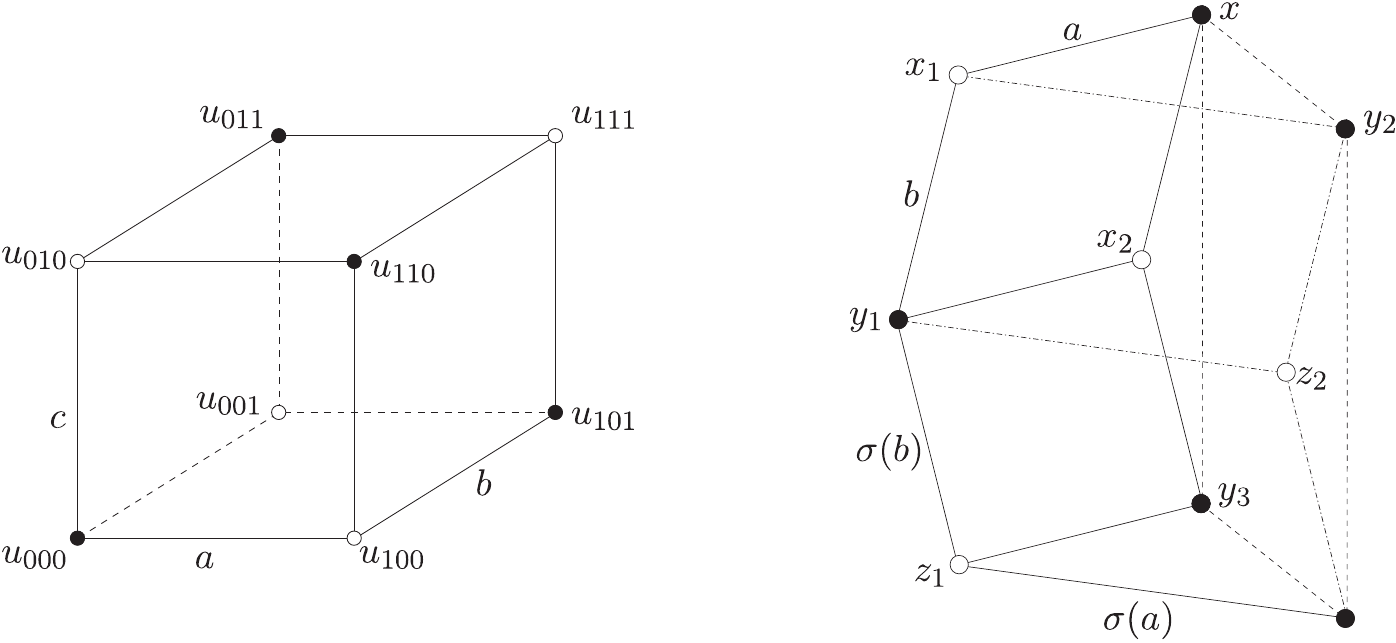}
\caption{3D consistency and 3D boundary consistency.
We recall that parallel edges carry the same parameter.}\label{fig:3DYBE}
\end{figure}

As explained in the introduction, we adopt the 3D consistency approach to integrable quad-graph equations.
Let us f\/irst recall what it means in the bulk case~\cite{BS2,Nij} before we introduce its boundary analog.
The usual setup is depicted in l.h.s.\ of Fig.~\ref{fig:3DYBE} where the bulk equation of motion $Q=0$ is
attached to each face of the cube.
Given values of the f\/ield in the three independent directions of the cube, say $u_{000}$, $u_{100}$,
$u_{010}$, $u_{001}$, there are three dif\/ferent ways of computing $u_{111}$ by repeated use of $Q=0$.
The 3D consistency condition requires that these three possibilities give the same result for $u_{111}$.

We propose, in the following, the main new equation of this article which is a~similar consistency
condition for the function $q$ (and $\sigma$) that we call the \textit{3D boundary consistency condition}.
This condition is in fact a~compatibility condition between the bulk equation $Q=0$ and the boundary
equation $q=0$.
Instead of the cube for the 3D consistency, the 3D boundary consistency lies on a~half of a~rhombic
dodecahedron as displayed in r.h.s.\ of Fig.~\ref{fig:3DYBE}.
Let us remark that this half of a~rhombic dodecahedron has the same combinatorial structure than the
r.h.s.\ of the Fig.~\ref{fig:trc}.
On each face (resp.~4 quadrilaterals and~4 triangles), we attach its corresponding equation of motion
(resp.\ $Q=0$ and $q=0$).
Then, the 3D boundary consistency is the statement that, given the values $x$, $x_1$ and $x_2$ of the
f\/ield, the dif\/ferent schemes to compute $w$ (see Fig.~\ref{fig:3DYBE}) lead to the same results.
More precisely, following the notations in Fig.~\ref{fig:3DYBE}, we get that, given $x$, $x_1$ and $x_2$:
\begin{itemize}\itemsep=0pt
\item the three equations
\begin{gather}
\label{scheme1}
Q(y_1,x_2,x_1,x;a,b)=0,
\qquad
q(x,x_1,y_2;a)=0,
\qquad
q(x,x_2,y_3;b)=0,
\end{gather}
gives respectively the values of $y_1$, $y_2$ and $y_3$.
\item Then, both equations
\begin{gather}
\label{scheme2}
Q(y_1,z_1,x_2,y_3;\sigma(b),a)=0,
\qquad
Q(y_1,z_2,x_1,y_2;\sigma(a),b)=0,
\end{gather}
gives respectively the values of $z_1$ and $z_2$.
\item Finally, the three equations
\begin{gather}
\label{scheme3}
q(y_2,z_2,w;b)=0,
\qquad
q(y_3,z_1,w;a)=0,
\qquad
Q(y_1,z_1,z_2,w;\sigma(b),\sigma(a))=0,
\end{gather}
provides three ways to compute $w$ which must give the same answer.
\end{itemize}
\begin{definition}
We say that we solve the 3D boundary consistency for $q$ if, given a~3D consistent~$Q$, we f\/ind
a~function $q$ of $x$, $y$, $z$, $a$ and a~function $\sigma$ of $a$ such that the
scheme~\eqref{scheme1}--\eqref{scheme3} gives a~unique value for $w$ given values for $x$,
$x_1$ and~$x_2$.
In this case, $q$ is called a~solution of the 3D boundary consistency condition (we omit explicit
reference to $\sigma$ which is taken as part of the solution).
We also say that~$q$ is compatible with~$Q$.
\end{definition}

Note that this should not be confused with the notion of a~solution of the actual dynamics described by $Q$
and $q$.
Such a~solution would consists in f\/inding an expression for the f\/ield $u$ at each vertex of the
quad-graph that satisf\/ies the bulk and boundary dynamics.
This is an exciting open question which deserves separate attention and is beyond the scope of the present
paper.

In the bulk case, when one f\/inds a~$Q$ which satisf\/ies the 3D consistency (see Fig.~\ref{fig:3DYBE}),
the associated system is called an integrable equation on quad-graph~\cite{ABS, BS2,Nij}.
To introduce a~boundary which preserves the integrability, for a~given $Q$, we must solve the 3D boundary
consistency condition, i.e.\ f\/ind compatible functions~$q$ and $\sigma$.
\begin{definition}
We call integrable equation on a~quad-graph with boundary the data of a~quad-graph with boundary with
compatible labelling as well as functions $Q$, $q$ and $\sigma$ which satisfy the 3D consistency and the 3D
boundary consistency conditions.
\end{definition}

We justify the adjective ``integrable'' in Sections~\ref{sec:bac} and~\ref{sec:ZCR} by showing the
presence of B\"acklund transformations and a~zero curvature representation.

Let us emphasize that similar approaches have already appeared in the literature to introduce integrable
boundaries in dif\/ferent contexts.
Indeed similar f\/igures to those in Fig.~\ref{fig:3DYBE} appeared in~\cite{AK,BPO,FHS} as the face
representation of the ref\/lection equation~\cite{Che,Sk}.
The right-hand-side of Fig.~\ref{fig:3DYBE} is also half of the f\/igure representing the tetrahedron
equation~\cite{Bax86,BMS}.
There exists also a~close connection between the set-theoretical equation introduced recently
in~\cite{CCZ, VZ2} and the 3D boundary consistency (see Section~\ref{sec:conn} for more details).
Similar connections have been studied previously in the bulk case where the set-theoretical Yang--Baxter
equation is linked to the 3D consistency condition~\cite{ABS,PTV}.

\section{Solutions of the 3D boundary consistency}
\label{sec:class}

In this section, we provide a~list of solutions of the 3D boundary consistency condition associated to the
bulk equations $Q$ classif\/ied in~\cite{ABS}.
In other words, given~$Q$, we want to f\/ind solutions $q$ and $\sigma$ which satisfy the~3D boundary
consistency conditions.
The underlying idea is that this should provide integrable boundary conditions for integrable discrete
equations characterized by~$Q$.

\subsection{The ABS classif\/ication}

For completeness, we list the solutions of the 3D consistency equations~\eqref{eq:Q} obtained
in~\cite{ABS}
\begin{alignat*}{3}
& {\rm (Q1)}:
\quad &&
a (u_{00}-u_{01})(u_{10}-u_{11})- b (u_{00}-u_{10})(u_{01}-u_{11}) + \delta^2 a b (a-b)= 0,&
\\
& {\rm (Q2)}:
\quad &&
a(u_{00}-u_{01})(u_{10}-u_{11})-b(u_{00}-u_{10})(u_{01}-u_{11}) &
\\
&&&
{} +a b(a-b)(u_{00}+u_{10}+u_{01}+u_{11}) - ab (a-b)\big(a^2-ab+b^2\big)=0, &
\\
& {\rm (Q3)}:
\quad &&
\big(b^2-a^2\big)(u_{00}u_{11}+u_{10}u_{01})+b\big(a^2-1\big)(u_{00}u_{10}+u_{01}u_{11}) &
\\
&&&
{}-a\big(b^2-1\big)(u_{00}u_{01}+u_{10}u_{11})-\delta^2\big(a^2-b^2\big)\big(a^2-1\big)\big(b^2-1\big)/(4ab)=0,&
\\
& {\rm (Q4)}:
\quad &&
\sn(a)(u_{00}u_{10}+u_{01}u_{11}) - \sn(b)(u_{00}u_{01}+u_{10}u_{11})-\sn(a-b)(u_{00}u_{11}+u_{10}u_{01}) &
\\
&&&
{}+\sn(a-b)\sn(a)\sn(b)(1+ K^2u_{00}u_{10}u_{01}u_{11})=0,
\\
& {\rm (H1)}:
\quad &&
(u_{00}-u_{11})(u_{10}-u_{01})+b-a=0, &
\\
& {\rm (H2)}:
\quad &&
(u_{00}-u_{11})(u_{10}-u_{01})+(b-a)(u_{00}+u_{10}+u_{01}+u_{11})+b^2-a^2=0, &
\\
& {\rm (H3)}:
\quad &&
a(u_{00}u_{10}+u_{01}u_{11})-b(u_{00}u_{01}+u_{10}u_{11})+\delta^2\big(a^2-b^2\big)=0, &
\\
& {\rm (A1)}:
\quad &&
a(u_{00}+u_{01})(u_{10}+u_{11})-b(u_{00}+u_{10})(u_{01}+u_{11})-\delta ^2ab(a-b)=0, &
\\
& {\rm (A2)}:
\quad &&
b\big(a^2-1\big)(u_{00}u_{01}+u_{10}u_{11}) -a\big(b^2-1\big)(u_{00}u_{10}+u_{01}u_{11})&
\\
&&&
{}+ \big(b^2-a^2\big)(u_{00}u_{10}u_{01}u_{11} +1)=0.&
\end{alignat*}
We use the same labels (Q, H, A families) and forms that were used in~\cite{ABS}, except for equation~(Q4)
which is in an equivalent form introduced in~\cite{AHN,Hie} where $\sn(a) = \sn(a; K)$ is the
Jacobi elliptic function with modulus~$K$.
It is worth noting that the 3D consistency condition as well as the af\/f\/ine-linearity, $D_4$-symmetry
and tetrahedron properties are preserved for each of the equations up to common M\"obius transformations on
the variables $u_{00}$, $u_{10}$, $u_{01}$, $u_{11} $ and point transformations on the parameters $a$, $b$.

\subsection{Method and solutions}
\label{Sec:method}
Instead of performing brute force calculations where one could make assumption on the form of $q$ (like
multilinearity in the variables) and insert in the 3D boundary consistency condition for a~given~$Q$,
below we describe a~simple method that amounts to ``fold'', in a~certain sense, $Q$ to obtain compatible
$q$'s.
Although it may seem ad hoc and arbitrary, this method has several motivations.
First, it gives a~simple way to obtain three-leg forms for~$q$ knowing the ones for~$Q$ and hence a~way to
introduce discrete Toda-type systems with boundary.
This is explained in Section~\ref{sec:toda}.
Second, the method is a~simple adaptation of the folding method that was used in~\cite{CCZ} to obtain
ref\/lection maps, i.e.\ solutions of the set-theoretical ref\/lection equation.
This last point is discussed in more detail in Section~\ref{sec:conn}.
There, we present an alternative method to f\/ind admissible $q$'s starting from ref\/lection maps.
This alternative method produces some of the solutions that are also obtained with the method that we
explain here.
But more importantly, this alternative method establishes a~deep connection between solutions of the 3D
boundary consistency condition and ref\/lection maps.
This is reminiscent of the deep connection between solutions $Q$ of the 3D consistency condition (in
particular the ones of the ABS classif\/ication) and quadrirational Yang--Baxter maps~\cite{ABS2}.

Now, given $Q$ satisfying the 3D consistency condition, we look for $q(u_{00},u_{10},u_{11};a)$ of the
form
\begin{gather}
\label{ansatz_q}
q(u_{00},u_{10},u_{11};a)=Q(u_{00},u_{10},k(u_{00},u_{10};a),u_{11};a,\sigma(a)),
\end{gather}
satisfying the 3D boundary consistency condition, where $k$ and $\sigma$ are the functions to be determined.
Equation~\eqref{ansatz_q} may be seen as the folding along the diagonal $(u_{00},u_{11})$ of the
quadrilateral in Fig.~\ref{quad_tri} to get the triangle $(u_{00},u_{10},u_{11})$.
Obviously, one may fold along the other diagonal but, due to the $D_4$-symmetry of $Q$, we get the same
results.

To f\/ind $k$ and $\sigma$ (and hence $q$), for each given $Q$, we insert our ansatz~\eqref{ansatz_q} in
the scheme~\mbox{\eqref{scheme1}--\eqref{scheme3}} and try to f\/ind functions $k$ and $\sigma$ that fulf\/ill the
resulting constraints.
We note that the following ``trivial'' choice
\begin{gather}
\label{eq:trivial}
\sigma(a)=a,
\qquad
k(u_{00},u_{10};a)=-u_{10},
\end{gather}
is always a~solution of the problem for any $Q$.
It yields $q(u_{00},u_{10},u_{11};a)=a u_{10}(u_{00}-u_{11})$.
We report in Tables~\ref{ta:Q81} and~\ref{ta:H81} the nontrivial functions $k$, $\sigma$ and $q$ we found
for the dif\/ferent $Q$'s of the~Q,~H and~A families of the ABS classif\/ication.
Note that we make no claim of completeness.
A~method for a~systematic classif\/ication is in fact an interesting open problem.
\begin{table}[t!]
\centering
\caption{Results for boundary equations (Q family).
$\mu$ is a~free parameter.
The asterisk denotes solu\-tions that are also obtained with the method of Section~\ref{sec:conn}.\label{ta:Q81}}

\vspace{2mm}

\begin{tabular}{@{}l@{\,\,\,\,}l@{\,\,\,\,}l@{\,\,\,\,}l@{}}
\hline
\bsep{2pt}\tsep{2pt} ABS & $\sigma(a)$ & $q(x,y,z;a)$ & $k(x,u;a)$
\\
\hline
\multirow{4}{*}{Q1$_{\delta=0}$ } & $\frac{\mu^2}{a}^*$ & $a (y-z)+\mu (x-y) $ & $ x+\frac{\mu}{a}(x-u)$\tsep{4pt}\bsep{2pt}
\\
& $\frac{\mu^2}{a}^*$ & $a x (y-z)+\mu z(y-x) $ &$\frac{a x u}{au+\mu(x-u)}$\bsep{2pt}
\\
& $-a+2\mu^*$ & $y(x+z) $ & $\frac{\mu x u+(a-\mu)x^2}{(a-\mu)u+\mu x}$\bsep{2pt}
\\
& $-a+2\mu^*$ & $a \left(y^2-x z\right)+(x-y) (y+z) \mu $ &$-u$\bsep{2pt}
\\
\hline
\multirow{6}{*}{ Q1$_{\delta=1}$} & $\frac{\mu^2}{a}^*$ & $a (y-z-\mu )\pm\mu (y-x-\mu )$ & $
x+\mu\pm\frac{\mu}{a}(u-x-\mu)$\tsep{4pt}\bsep{2pt}
\\
& $-a+2\mu^*$ & $y(x-z)$ &$ x+\frac{a(a-2\mu)}{x-u}$\bsep{2pt}
\\
& $-a+2\mu^*$ & $(y-x) (y-z)+a(a-2\mu) $ &$u$\bsep{2pt}
\\
& $-a+2\mu$ & $y(x+z) $ & $ \frac{x(ax-\mu(x-u))-a(a-\mu)(a-2\mu)}{au+\mu(x-u)}$\bsep{2pt}
\\
& $-a+2\mu$ & $(y^2-x z)+\frac{\mu}{a}(x-y)(y+z)$   &\multirow{2}{*}{$-u$}\bsep{2pt}\\
& & $\quad{}-(a-\mu)(a-2\mu)$ &\bsep{2pt}
\\
\hline
\multirow{4}{*}{Q3$_{\delta=0}$ } & $\pm\frac{\mu}{a}^*$ & $y(x\pm z)$ & $\pm\frac{((a^2-\mu)x-a(1-\mu)u
)x}{(a^2-\mu)u-a(1-\mu)x}$\tsep{4pt}\bsep{2pt}
\\
& $\pm \frac{\mu}{a}^*$ & $(a^2+ \mu )\left(y^2\pm x z\right) $ &\multirow{2}{*}{$\pm u$}\bsep{2pt}
\\
&&$\quad{}-a(1+ \mu ) y (x\pm z)$ & \bsep{2pt}\\
& $- a$ & $y(x+ z)$ & $-u$\bsep{2pt}
\\
\hline
\multirow{5}{*}{Q3$_{\delta=1}$ } & $\frac{\mu}{a}$ & $y(x\pm z)$ & $\frac{a(\mu\mp 1)xu-(\mu\mp
a^2)x^2+\frac{1}{4}(a^2-1)(a^2-\mu^2)\left(\frac{1}{\mu}\mp\frac{1}{a^2}\right)}{(a^2\mp\mu)u-a(1\mp \mu)x}$\tsep{6pt}\bsep{2pt}
\\
& \multirow{2}{*}{$\frac{\mu}{a}$} & $a(\mu\pm 1) y (x \pm z)$ &
\multirow{3}{*}{$\pm u$}\bsep{2pt}
\\
&&$\quad{}-(\mu\pm a^2)(y^2\pm xz)$ & \bsep{2pt}
\\
& &  $\quad{}+\frac{1}{4}(\frac{1}{\mu}\pm\frac{1}{a^2})(a^2-1)(a^2-\mu^2)$ &\bsep{2pt}
\\
& $-a$ & $y(x+ z)$ &$-u$\bsep{2pt}
\\
\hline
\multirow{2}{*}{ Q4} & $-a$ & $y(x+z)$ &$ \frac{x^2-\sn^2(a)}{u(1-K^2\sn^2(a)x^2)}$\tsep{4pt}\bsep{2pt}
\\
& $-a$ &$\sn^2(a)(K^2 y^2xz-1)+y^2-xz$ & $-u$\bsep{2pt}
\\
\hline
\end{tabular}
\end{table}

\begin{table}[t!]\centering
\caption{Results for boundary equations (H and A families).
$\mu$ is a~free parameter.
The asterisk denotes solutions that are also obtained with the method of Section~\ref{sec:conn}.\label{ta:H81}}
\vspace{2mm}

\begin{tabular}{@{}l@{\,\,\,}l@{\,\,\,}l@{\,\,\,}l@{}}
\hline
\\
ABS & $\sigma(a)$ & $q(x,y,z;a)$ & $k(x,u;a)$\tsep{2pt}\bsep{2pt}
\\
\hline
\multirow{2}{*}{H1} & $-a+2\mu^*$ & $y (x+z)$ & $ u+\frac{\mu-a}{x}$\tsep{2pt}\bsep{2pt}
\\
& $-a+2\mu^*$ & $y(z-x)+a-\mu$ &$-u$\bsep{2pt}
\\
\hline
\multirow{4}{*}{H2 } & $-a+\mu^*$ & $x+2 y+z+\mu $ & $u$\tsep{2pt}\bsep{2pt}
\\
& $-a+\mu^*$ & $y(z-x)$ & $-2x-u-\mu$\bsep{2pt}
\\
& $-a+\mu$ & $2y(x-z)+(\mu-2a)(x+z+\mu)$ & $-u$\bsep{2pt}
\\
& $-a+\mu$ & $y(x+z)$ & $ \frac{2ux+(u+\mu)(\mu-2a)}{2x+2a-\mu}$\bsep{2pt}
\\
\hline
\multirow{2}{*}{H3$_{\delta=0}$} & $-a$ & $y(x+z)$ & $-u$\tsep{2pt}\bsep{2pt}
\\
& $\frac{\mu}{a}$ & $y(x\pm z)$ & $\pm u$\bsep{2pt}
\\
\hline
\multirow{3}{*}{H3$_{\delta=1}$} & $\frac{\mu}{a}^*$ & $y(x\pm z)$ & $  \pm u-\frac{\mu \mp
a^2}{ax}$\tsep{6pt}\bsep{2pt}
\\
& $\frac{\mu}{a}^*$ & $a^2+a y (x\pm z)\pm \mu $ & $\pm u$\bsep{2pt}
\\
& $-a$ & $y(x+z)$ & $-u$\bsep{2pt}
\\
\hline
\multirow{4}{*}{A1$_{\delta=0}$ } & $\frac{\mu^2}{a}^*$ & $\mu(x+y)+a(y+z)$ & $-x+\frac{\mu}{a}(u+x)$\tsep{6pt}\bsep{2pt}
\\
& $\frac{\mu^2}{a}^*$ & $a x (y+z)+(x+y) z \mu $ &$\frac{axu}{\mu(u+x)-au}$\bsep{2pt}
\\
& $-a+2 \mu^*$ & $y(x+z)$ & $\frac{(ax-\mu(u+x))x}{au-\mu(u+x)}$\bsep{2pt}
\\
& $-a+2 \mu^*$ & $a \left(y^2-x z\right)-(x+y) (y-z) \mu $ &$-u$\bsep{2pt}
\\
\hline
\multirow{5}{*}{A1$_{\delta=1}$} & $\frac{\mu^2}{a}^*$ & $a (y+z-\mu )\pm \mu (x+y-\mu ) $ &$-x+\mu\pm
\frac{\mu}{a}(x+u-\mu)$\tsep{2pt}\bsep{2pt}
\\
& $-a+2 \mu^*$ & $y(z-x) $ &$-x-\frac{a(a-2\mu)}{x+u}$\bsep{2pt}
\\
& $-a+2 \mu^*$ & $(x+y) (y+z) +a(a-2\mu)$ &$u$\bsep{2pt}
\\
& $-a+2 \mu$ & $y(x+z) $ &$\frac{x(ax-\mu(x+u))-a(a-\mu)(a-2\mu)}{au-\mu(x+u)}$\bsep{2pt}
\\
& $-a+2 \mu$ & $\frac{\mu}{a} (x+y) (y-z) -\left(y^2-x z\right)+(a-\mu)(a-2\mu) $ &$-u$\bsep{2pt}
\\
\hline
\multirow{3}{*}{A2} & $\pm \frac{\mu}{a}^*$ & $y(z\pm x)$ &$\pm
\frac{a(\mu-1)ux+a^2-\mu}{(a^2-\mu)xu+a(\mu-1)} $\tsep{6pt}\bsep{2pt}
\\
& $\pm \frac{\mu}{a}^*$ & $a (1+\mu ) y(x\pm z) -(a^2+\mu) \left(1\pm x y^2 z\right) $ &$\pm u$\bsep{2pt}
\\
& $-a$ & $y(z+ x)$ &$-u$\bsep{2pt}
\\
\hline
\end{tabular}
\end{table}

\section{Other aspects of integrable equations on quad-graphs\\ with boundary}
\label{sec:aspects}

In this section, we present results on important traditional aspects of integrability obtained from the 3D
boundary consistency equation.
They justify a~posteriori our def\/inition of integrability for quad-graphs with boundary.

\subsection{B\"acklund transformations}
\label{sec:bac}

In this subsection, we prove that the 3D boundary consistency condition proposed in the previous section
leads naturally to B\"acklund transformations which are a~basic tool in the context of classical
integrability and soliton theory.
This result is similar to the one without boundary~\cite{ABS} and is summarized in the following
proposition:
\begin{proposition}
Let us suppose that we have an integrable equation on a~quad-graph with boundary $($with the set of all its
vertices denoted as $V)$ as well as a~solution, $g:V\rightarrow {\mathbb C}$.
There exist a~two-parameter solution $g^+$ of the same integrable quad-graph equation and a~function~$f$
from~$V$ to~${\mathbb C}$ satisfying
\begin{gather*}
Q(g(v),g(v_1),f(v),f(v_1);a,\lambda)=0,
\qquad
Q\big(f(v),f(v_1),g^+(v),g^+(v_1);a,\sigma(\lambda)\big)=0,
\end{gather*}
for all edges $(v,v_1)$ of the quad-graph not on the boundary of the surface $(a$ is the parameter
associated to this edge$)$, and
\begin{gather*}
q\big(g(v_2),f(v_2),g^+(v_2);\lambda\big)=0,
\end{gather*}
for all vertices $v_2$ on the boundary of the quad-graph.
We call the solution $g^+$ the B\"acklund transform of $g$.
\end{proposition}
\begin{proof}
The proof follows the same lines as in the case without boundary.
We start with the quad-graph with boundary, called in this proof the ground f\/loor.
We consider also two other copies of the surface, called f\/irst and second f\/loor, of the ground f\/loor.
Then, we construct a~3D graph by the following procedure (see also Fig.~\ref{fig:back}):
\begin{itemize}
\itemsep=0pt \item There is a~one-to-one correspondence between the vertices of the ground f\/loor and the
ones of the f\/irst f\/loor but one moved the vertices of the f\/irst f\/loor such that no vertex of the
f\/irst f\/loor lies on the boundary of the surface (see l.h.s.\ of the Fig.~\ref{fig:back}).
The vertices on the second f\/loor is an exact copy of the ones on the ground f\/loor.
\item We copy the edges in the bulk of the ground f\/loor on the f\/irst and second f\/loor.
The copies carry the same label.
We copy the edges on the boundary of the ground f\/loor only on the second f\/loor.
\item We add the edges (thin lines on the Fig.~\ref{fig:back}) linking all the vertices of the ground
f\/loor with the corresponding vertices of the f\/irst f\/loor and similarly from the f\/irst to the second
f\/loor.
The edges between the ground and f\/irst f\/loor carry the label $\lambda$ whereas the edges between the
f\/irst and second f\/loor carry $\sigma(\lambda)$.
We add also the edges between the vertices on the boundary of the ground f\/loor to the corresponding ones
on the second f\/loor.
\item the set of faces is the union of the following sets: (i) the triangular and quadrilateral faces of
the ground and second f\/loor; (ii) the quadrilateral faces of the f\/irst f\/loor; (iii) the ``vertical''
quadrilateral faces made from the edges of the ground f\/loor, the corresponding ones of the f\/irst
f\/loor and the vertical edges linking the vertices of these edges, and similarly between the f\/irst and
second f\/loors; (iv) the ``vertical'' triangular faces made of the edges linking the vertices on the
boundary of the ground and second f\/loors and the corresponding vertex of the f\/irst f\/loor (which is
not on the boundary).
\end{itemize}
\begin{figure}[t]
\centering
\includegraphics{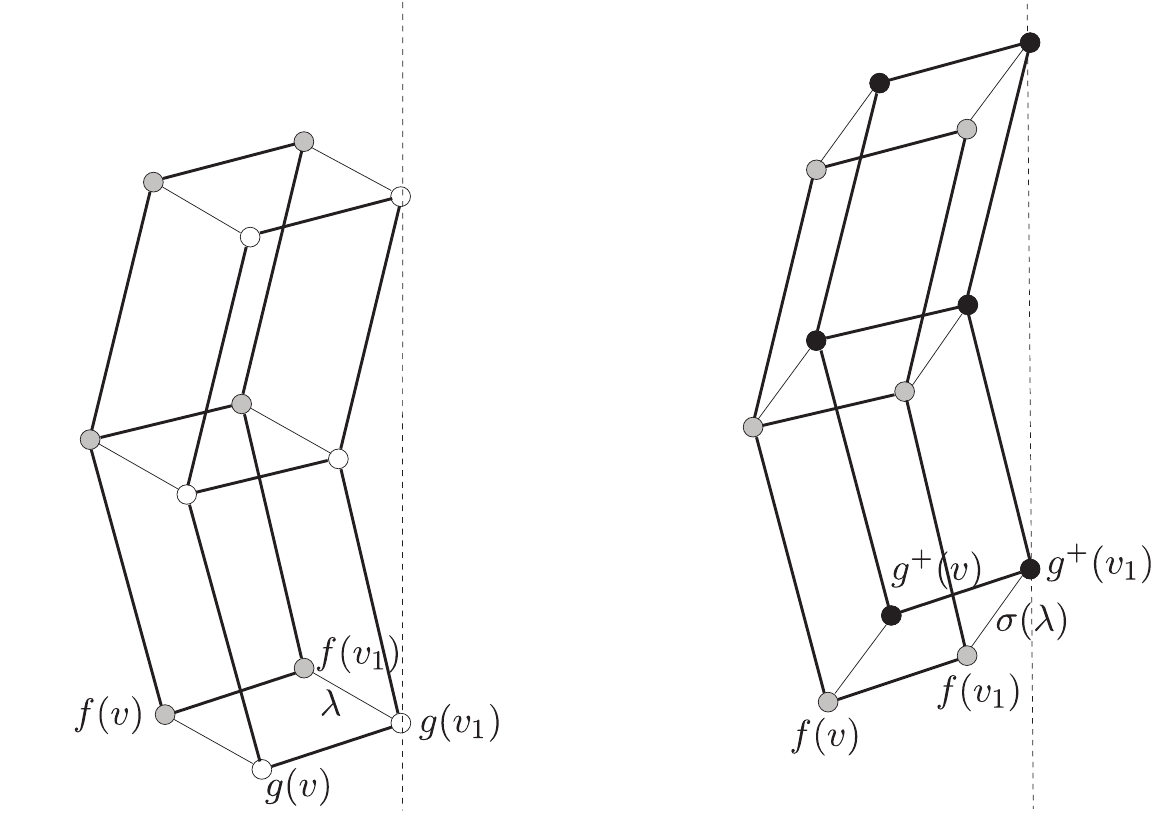} \caption{The f\/igure on the left shows the ground f\/loor
(white vertices and thick lines connecting them) and the f\/irst f\/loor (gray vertices and thick lines
connecting them) corresponding to a~typical conf\/iguration near the boundary.
The f\/irst f\/loor and the second one (black vertices and thick lines connecting them) is shown on the
f\/igure on the right.
The thin lines correspond to the edges linking the dif\/ferent f\/loors: they carry~$\lambda$ on the left
and~$\sigma(\lambda)$ on the right.}\label{fig:back}
\end{figure}

We impose now that the values of the f\/ield living on this 3D graph be constrained by $Q=0$ on each
quadrilateral face and by $q=0$ on each triangular faces.

As in the case without boundary, due to the 3D consistency condition, given a~function $g$ satisfying the
constraints on the ground f\/loor, one can get a~function $f$ satisfying the constraint on the f\/irst
f\/loor depending on $\lambda$ and on the value of $f$ at one vertex of the f\/irst f\/loor.
This function $f$ satisf\/ies, in particular, $Q(g(v),g(v_1),f(v),f(v_1);a,\lambda)=0$.

Knowing the value of the f\/ield at one vertex on the boundary of the ground f\/loor (say~$g(v_1)$ on the
f\/igure) and the corresponding value on the f\/irst f\/loor~($f(v_1)$), we get the value~$g^+(v_1)$ of the
f\/ield~$g^+$ at the corresponding vertex of the second f\/loor using the equation $q(g(v_1), f(v_1),
g^+(v_1)$, $\lambda)=0$ which connects the white, grey and black copies of the vertex~$v_1$.

We obtain a~function $g^+$ satisfying the constraints of the second f\/loor using $Q$ on the ``vertical''
quadrilateral faces between the f\/irst and second f\/loors.
The important point is to remark that due to the 3D consistency and the 3D boundary consistency
conditions, all the dif\/ferent ways to obtain the values of $g^+$ give the same result.
We may see this geometrically since the only ``elementary blocks'' of the 3D graph are cubes or half of
rhombic dodecahedron and since the functions $Q$ and $q$ are chosen such that they satisfy the 3D
consistency and the 3D boundary consistency conditions, the dif\/ferent ways to compute the values of the
f\/ield $g^+$ are consistent.

Finally, using the fact that the second f\/loor is an exact copy of the ground f\/loor, $g^+$ satisf\/ies
the constraints of the original quad-graph equations.
\end{proof}

Let us remark that the main dif\/ference between the cases with and without boundary lies in the necessity
of an additional, intermediate f\/loor in the case with boundary.
This feature appeared previously in the context of asymmetric quad-graph equations~\cite{Boll}.
Let us emphasize that the function $f$ def\/ined on the intermediate f\/loor is not in general a~solution
of the same equations on a~quad-graph with boundary since the values of the f\/ield on the vertices of the
``would-be'' boundary of the f\/irst f\/loor do not necessarily satisfy an equation of the
type~\eqref{eq:q}.

\subsection{Zero curvature representation}
\label{sec:ZCR}

It is well established that a~3D consistent system $Q(u_{00},u_{10},u_{01},u_{11};a,b)$ admits a~zero
curvature representation~\cite{BS2,BS,Nij}, i.e.\ there exists a~matrix $L$ depending on values of the
f\/ield on the same edge, the parameter associated to this edge and a~spectral parameter $\lambda$ such
that the following equation holds\footnote{This equation holds projectively but a~choice of the
normalization of $L$ allows us to transform it into a~usual equality.}
\begin{gather}
\label{eq:ZCR}
L(u_{11},u_{10},b;\lambda)~L(u_{10},u_{00},a;\lambda)~=~L(u_{11},u_{01},a;\lambda)~L(u_{01},u_{00}
,b;\lambda).
\end{gather}

There exists a~constructive way to get $L$ from $Q$~\cite{BS2,Nij}: due to the linearity of the
function~$Q$, we can rewrite equivalently equation~\eqref{eq:Q} as follows
\begin{gather}
\label{eq:QL}
u_{11}=L(u_{10},u_{00},a;b)[u_{01}],
\end{gather}
where $L$ is a~2 by 2 matrix describing, with usual notations, a~M\"obius transformation
\begin{gather*}
L[z]=\frac{\alpha z+\beta}{\gamma z+\delta}
\qquad
\text{where}
\qquad
L=\left(
\begin{matrix}
\alpha&\beta
\\
\gamma&\delta
\end{matrix}
\right).
\end{gather*}
Geometrically, using Fig.~\ref{fig:3DYBE}, it is easy to show that the matrix~\eqref{eq:QL} satisf\/ies the
zero curvature equation~\eqref{eq:ZCR} if $Q$ satisf\/ies the 3D consistency~\cite{BS2,BS,Nij}.

Similarly, we want to show that the 3D boundary consistent system admits a~zero curvature representation.
For the boundary equation $q(x,y,z;a)=0$, we propose the following zero curvature representation
\begin{gather}
\label{eq:ZCRb}
K(z;c)L(z,y,\sigma(a);c)L(y,x,a;c)=L(z,y,\sigma(a);\sigma(c))L(y,x,a;\sigma(c))K(x;c),
\end{gather}
where $K$ is also a~2 by 2 matrix.
We can now state the following results justifying the previous def\/inition:
\begin{proposition}
All the examples of boundary equations $q(x,y,z;a)=0$ displayed in Tables~{\rm \ref{ta:Q81}} and~{\rm \ref{ta:H81}} as
well as the trivial solution~\eqref{eq:trivial} can be represented by equation~\eqref{eq:ZCRb} where
$K(x;a)[u]$ is the~$2$ by~$2$ matrix describing the function $k(x,u;a)$, understood as a~M\"obius
transformation w.r.t.\
$u$ with polynomial coefficients in~$x$ and~$a$.
\end{proposition}
\begin{proof}
We give the details of the proof for the case given in the f\/irst row of the Table~\ref{ta:Q81}, i.e.\ we
deal with the bulk equation ${\rm Q1}_{\delta=0}$ given by
\begin{gather*}
Q(u_{00},u_{10},u_{01},u_{11};a,b)=a(u_{00}-u_{01})(u_{10}-u_{11})-b(u_{00}-u_{10})(u_{01}-u_{11}),
\end{gather*}
and the boundary equation characterized by
\begin{gather}
q(x,y,z;a)=a(y-z)+\mu(x-y)
\qquad
\text{with}
\quad
\sigma(a)=\frac{\mu^2}{a},
\nonumber
\\
k(x,u;a)=\frac{-\mu u+x(\mu+a)}{a}.
\label{eq:kp}
\end{gather}
The matrix $L$ associated to this $Q$ is given by
\begin{gather*}
L(y,x,a;b)=\frac{1}{\sqrt{b}\sqrt{b-a}(x-y)}\left(
\begin{matrix}
ay+b(x-y)& -axy
\\
a& -ax+b(x-y)
\end{matrix}
\right).
\end{gather*}
It is a~known result that equation~\eqref{eq:ZCR} with this choice for $L$ is satisf\/ied if
$Q(u_{00},u_{10},u_{01},u_{11};a,b)$ $=0$
but it is easily verif\/ied.
Let us mention that the parameters entering in the square roots of the normalisation of $L$ may be negative.
Therefore, one must choose a~branch cut for the square root appearing in the normalisation: we choose the
half-line $\{ix\,|\,x<0\}$.

The matrix $K$ associated to the function $k$ given by~\eqref{eq:kp} is
\begin{gather*}
K(x;a)=\left(
\begin{matrix}
-1 &\big(\frac{a}{\mu}+1\big)x
\\
0 &\frac{a}{\mu}
\end{matrix}
\right).
\end{gather*}
By algebraic computation, one gets
\begin{gather*}
K(z;c)~L(z,y,\sigma(a);c)~L(y,x,a;c)-L(z,y,\sigma(a);\sigma(c))~L(y,x,a;\sigma(c))~K(x;c)
\\
\qquad
\propto q(x,y,z;a)\left(
\begin{matrix}
a(y-z) & -ax(y-z)+cz(y-x)
\\
0 & c(y-x)
\end{matrix}
\right).
\end{gather*}
Therefore, if $q(x,y,z;a)=0$, relation~\eqref{eq:ZCRb} holds.
That proves the proposition for this case.
All the other cases are treated similarly which f\/inishes the proof of the proposition.
\end{proof}

Note that equation~\eqref{eq:ZCRb} provides a~rather general framework for the representation of an
integrable boundary in quad-graph models.
In the next section, we show how it contains as a~particular case a~previous approach to boundary
conditions in fully discrete systems.

\section{Toda-type models}
\label{sec:toda}

\subsection{Three-leg form}

It is known that any quad-graph equation $Q(x,u,v,y;a,b)=0$ of the ABS classif\/ication can be written
equivalently in the so-called \textit{three-leg form}~\cite{BS2}, either in an additive form,
\begin{gather}
\label{eq:Q31}
\psi(x,u;a)-\psi(x,v;b)=\phi(x,y;a,b),
\end{gather}
or a~multiplicative form,
\begin{gather}
\label{eq:Q32}
\psi(x,u;a)/\psi(x,v;b)=\phi(x,y;a,b).
\end{gather}

As demonstrated in~\cite{BS2,BS}, the existence of a~three-leg form leads to discrete systems of
Toda-type~\cite{Adl}.
Indeed, let $x$ be a~common vertex of the $n$ faces $(x,y_k,x_k,y_{k+1})$ (with $k=1,2,\dots,n$ and
$y_{n+1}=y_1$) with the parameters $a_k$ assigned to the edge $(x,y_k)$.
On each face, there is the equation $Q(x,y_k,y_{k+1},x_k;a_k,a_{k+1})=0$ written in the
presentation~\eqref{eq:Q31} or~\eqref{eq:Q32}.
Then, summing the corresponding $n$ equations of type~\eqref{eq:Q31}, one gets
\begin{gather*}
\sum_{k=1}^n\phi(x,x_k;a_k,a_{k+1})=0,
\end{gather*}
where $a_{n+1}=a_1$.
Similarly, multiplying $n$ equations of the type~\eqref{eq:Q32} leads to
\begin{gather*}
\prod_{k=1}^n\phi(x,x_k;a_k,a_{k+1})=1,
\end{gather*}
where $a_{n+1}=a_1$.
When the graph is bi-partite (for example with black and white vertices), we can reproduce this procedure
by taking as common vertices all the black vertices to get a~Toda-type model on the black subgraph.

\begin{figure}[t]\centering
\includegraphics{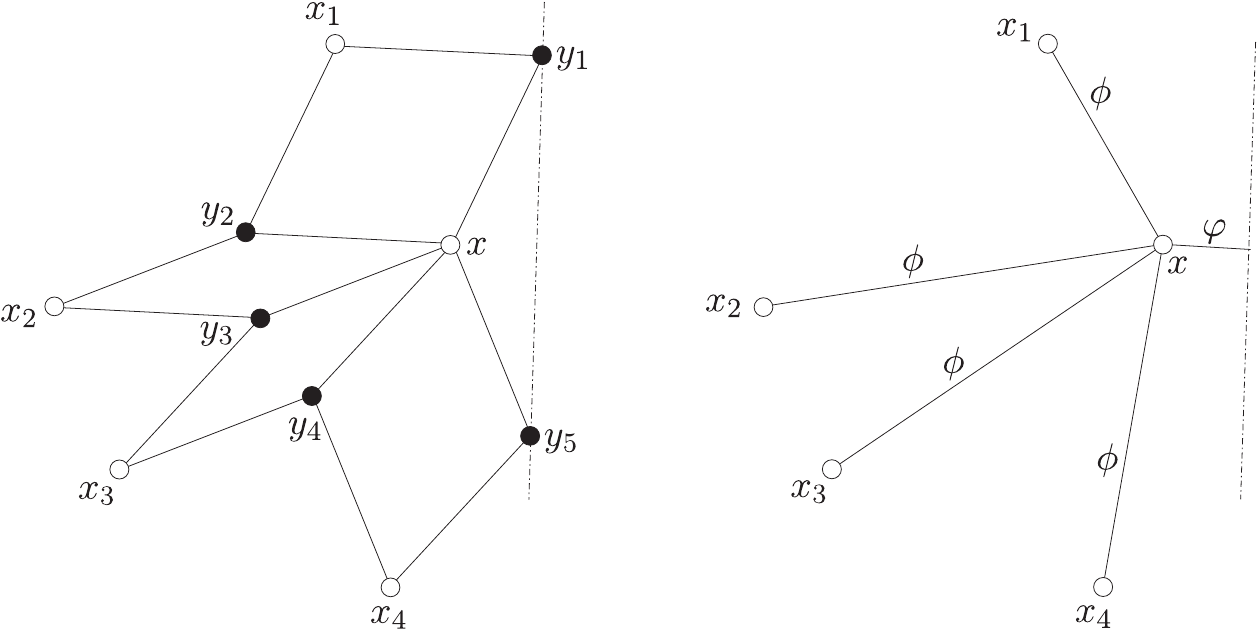} \caption{Faces adjacent to the vertex $x$ near a~boundary
(left).
Star graph from the white subgraph where the edges carry the functions associated appearing in the
Toda-type equation (right).}\label{fig:toda}
\end{figure}

Now assume that the boundary equation $q(x,y,z;a)=0$ can be written as
\begin{gather}
\label{eq:2leg}
\psi(y,x;a)-\psi(y,z;\sigma(a))=\varphi(y;a)\qquad \text{or}
\qquad \psi(y,x;a)/\psi(y,z;\sigma(a))=\varphi(y;a),
\end{gather}
where the function $\psi$ is the same as in the bulk case and the new function $\varphi$ depends only on
the central vertex of the triangle representing the boundary and on the parameter $a$.
In this case, one can obtain systems of Toda-type with boundary.
Indeed, let $x$ be a~vertex close to the boundary (i.e.
belonging to a~triangle but not sitting on the boundary) and common to $n-1$ quadrilateral faces
$(x,y_k,x_k,y_{k+1})$ (with $k=1,2,\dots,n-1$).
As in the bulk case, on each quadrilateral face, there is the equation $Q(x,y_k,y_{k+1},x_k;a_k,a_{k+1})=0$
written in the presentation~\eqref{eq:Q31} or~\eqref{eq:Q32}.
On the triangular face, it holds that $q(y_n,x,y_{1};a_n)=0$ (with $a_1=\sigma(a_n)$).
Then, summing (or multiplying) the corresponding $n$ equations, we get in the additive case
\begin{gather*}
\varphi(x;a_{n})+\sum_{k=1}^{n-1}\phi(x,x_k;a_k,a_{k+1})=0,
\end{gather*}
and in the multiplicative case
\begin{gather*}
\varphi(x;a_{n})\prod_{k=1}^{n-1}\phi(x,x_k;a_k,a_{k+1})=1.
\end{gather*}
We illustrate this procedure schematically in Fig.~\ref{fig:toda} for $n=4$.

As in the bulk case, if the graph is bi-partite (see footnote~\ref{footnote_bi}) and moreover if the
vertices on the boundary are of the same type, one can reproduce the above procedure on the whole graph to
get a~Toda-type model on a~subgraph and with a~boundary determined by~$\varphi$.
The conditions on the graph seem, at f\/irst glance, very restricting.
Nevertheless, the procedure explained in Section~\ref{sec:qgb} provides, starting from any graph,
examples satisfying such constraints.

\subsection{Boundary conditions for Toda-type systems}

In general, it seems that not all the solutions for $q$ found previously can be written as
in~\eqref{eq:2leg}.
However, it turns out that for each solution $q(x,y,z;a)$ for which the corresponding function $k(x,y;a)$
depends only on~$y$ or the function $q(x,y,z;a)$ factorizes as $f(y,a)g(x,z,a)$, there is a~way to
obtain~$\varphi$ from~$\phi$.
This is reminiscent, at the three-leg form level, of the simple folding procedure~\eqref{ansatz_q} used to
obtain~$q$ from~$Q$.
In these cases, the function $k(x,y;a)$ plays the role of the cut-of\/f constraint (or boundary condition)
for the Toda-type model in the sense of~\cite{HK} (see Example~\ref{ex:Ha} below for a~precise connection).

For the f\/irst case where the function $k(x,u;a)=k(u;a)$ does not depend on $x$, using
equation~\eqref{ansatz_q} in~\eqref{eq:Q31} or~\eqref{eq:Q32} together with the $D_4$-symmetry property
for~$Q$, one can see that equation~\eqref{eq:2leg} with the following function $\varphi$
\begin{gather}
\label{eq:f3leg}
\varphi(y;a)=\phi(y,k(y;a);a,\sigma(a)),
\end{gather}
is equivalent to the corresponding boundary equation $q(x,y,z;a)=0$.
Therefore, by using the explicit forms of $\phi$ given in~\cite{ABS} associated to each $Q$ of the ABS
classif\/ication and the results of Tables~\ref{ta:Q81} and~\ref{ta:H81}, we can derive $\varphi$ and hence
three-leg forms of the boundary equation in the form~\eqref{eq:2leg}.
In turn, this allows us to def\/ine Toda-type systems with a~boundary.
\begin{example}
The trivial solution~\eqref{eq:trivial} always corresponds to $\varphi(y;a)=0$ for the additive case or
$\varphi(y;a)=1$ for the multiplicative case.
This boundary condition may be interpreted as a~free boundary for the corresponding Toda-type system.
\end{example}
\begin{example}
We recall that $\psi(x,u;a)=\frac{a}{x-u}$ and $\phi(x,y;a,b)=\psi(x,y;a-b)$ for ${\rm Q1}_{\delta=0}$.
Looking at the fourth solution for ${\rm Q1}_{\delta=0}$ of Table~\ref{ta:Q81}, i.e.\ $\sigma(a)=-a+2\mu$ and
$k(x,y;a)=-y$, we obtain~$\varphi$ using~\eqref{eq:f3leg} and the following discrete Toda-type system with
a~boundary term:
\begin{gather*}
\frac{a_n-\mu}{x}+\sum_{k=1}^{n-1}\frac{a_k-a_{k+1}}{x-x_k}=0.
\end{gather*}
\end{example}

\begin{figure}[t]
\begin{minipage}[b]{75mm}\centering
\includegraphics[height=68mm]{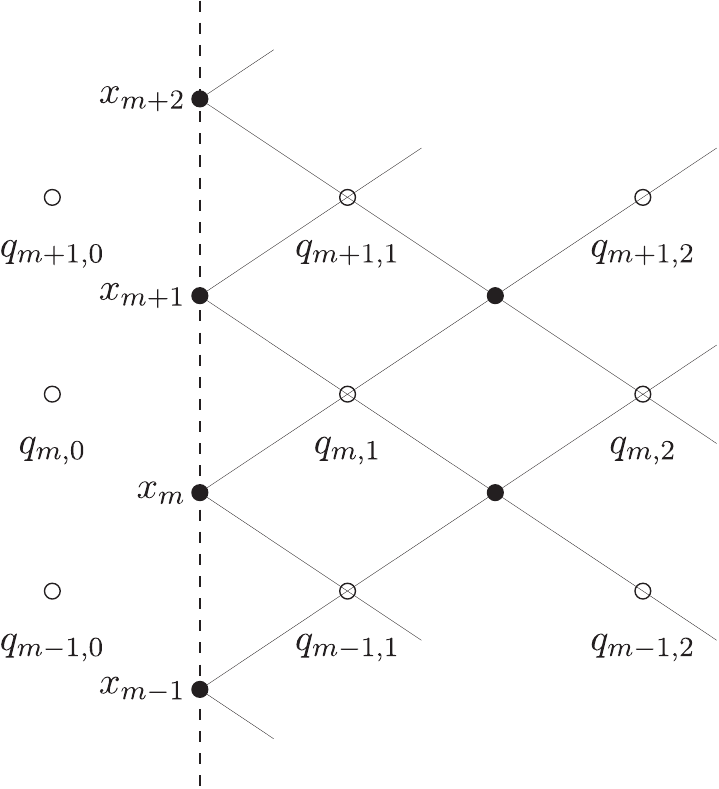}
\caption{A quad-graph with boundary with an underlying ${\mathbb Z}^2$ lattice structure. The additional white vertices will play the role
    of boundary vertices for the Toda-type system with boundary li\-ving on the white sublattice.}
  \label{fig:boundaryHAB1}
\end{minipage}
\hspace{8mm}
\begin{minipage}[b]{75mm}\centering
\includegraphics[height=68mm]{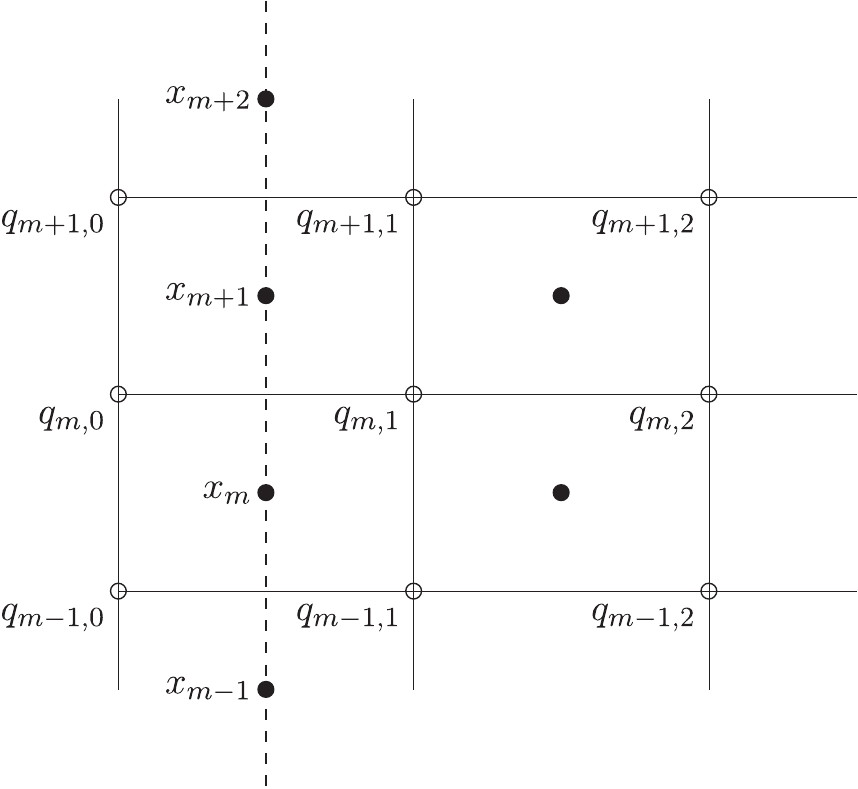}
\caption{The (white) ${\mathbb Z}^2$ (sub)lattice sup\-por\-ting the Toda-type system with a boundary. Our boundary is inserted between the boundary
    for the Toda-type system at site $n=0$ and the first site of the bulk system at $n=1$.}
    \label{fig:boundaryHAB}
\end{minipage}

\end{figure}

For the second case where the function $q(x,y,z;a)$ factorises, the construction is a~bit more involved.
The equation $q(x,y,z;a)=0$ constrains $x$ and $z$ independently of the values of $y$.
Therefore, we get equations involving only the values of the f\/ield on the boundary.
Let us suppose we solve these constraints on the boundary and denote by $\bar x$ the corresponding solution
for the f\/ield $x$.
These values on the boundary play the role of parameters in the function $k(\bar x,y;a)$ appearing in the
boundary conditions.
Also, one can see that equation~\eqref{eq:2leg} involves the following function $\varphi_{\bar x}$
\begin{gather*}
\varphi_{\bar x}(y;a)=\phi\big(y,k(\bar x,y;a);a,\sigma(a)\big),
\end{gather*}
and is equivalent to the corresponding boundary equation $q(\bar x,y,\bar z;a)=0$.
It also appears in the corresponding Toda-type system with boundary.
We now illustrate this case.
\begin{example}
\label{ex:Ha}
Let us restrict our general framework to a~${\mathbb Z}^2$ lattice system: we consider the quad-graph with
boundary represented on Fig.~\ref{fig:boundaryHAB1} with the bulk equations given by ${\rm Q1}_{\delta=0}$ with
the label $a$ on the lines $m-n={\rm const}$ and the label $b\neq a$ on the lines $m+n={\rm const}$.
The corresponding Toda-type model reads, for $n\ge 2$,
\begin{gather}
\label{eq3_Habi}
\frac{1}{q_{m,n+1}-q_{m,n}}-\frac{1}{q_{m,n}-q_{m,n-1}}=\frac{1}{q_{m+1,n}-q_{m,n}}-\frac{1}{q_{m,n}
-q_{m-1,n}},
\end{gather}
We use the third solution for $q$ for ${\rm Q1}_{\delta=0}$ in Table~\ref{ta:Q81} to generate the boundary
condition on the Toda-type system from the following boundary equation\footnote{We suppose that $q_{m,1}$
does not vanish.}
\begin{gather*}
q(\bar x_m,q_{m,1},\bar x_{m+1};a)=0
\qquad
\Leftrightarrow
\qquad
\bar x_m+\bar x_{m+1}=0,
\end{gather*}
on the quad-graph with boundary.
It is obvious that the general solution for the boundary values of the f\/ield is $\bar x_{m}=(-1)^m
\bar x_0$ (for any $\bar x_0$).
Then, following the procedure of folding explained in Section~\ref{Sec:method}, we obtain
\begin{gather}
\label{eq:BCtoda}
q_{m,0}=k(\bar x_m,q_{m,1};a),
\end{gather}
where
\begin{gather*}
k(\bar x_m,q_{m,1};a)=\frac{\mu(-1)^m\bar x_0q_{m,1}+(a-\mu)\bar x_0^2}{(a-\mu)q_{m,1}+\mu(-1)^m\bar x_0}.
\end{gather*}
In particular, we recover exactly the results of~\cite{HK} (equations~(13) and~(17)) for the
model~\eqref{eq3_Habi} (equation~(3) of~\cite{HK}) with the identif\/ications
\begin{gather}
\label{id}
\bar x_0\rightarrow\sqrt{\frac{a}{b}},
\qquad
a\rightarrow b+c\sqrt{\frac{b}{a}}
\qquad
\text{and}
\qquad
\mu\rightarrow c\sqrt{\frac{b}{a}},
\end{gather}
\end{example}

\subsection{Lax presentation of boundary conditions for a~Toda-type model}

In the previous Example~\ref{ex:Ha}, we showed that we can recover the boundary conditions introduced
in~\cite{HK}.
In this subsection, we show that the correspondence goes beyond this and is in fact also valid at the level
of the zero curvature representation.
Indeed, our zero curvature representation~\eqref{eq:ZCRb} allows us to recover as a~particular case the
main equation of~\cite{HK}, which we reproduced as~\eqref{boundary_Hab} below, and which encodes the
symmetry approach to integrable boundary conditions applied to integrable discrete chains.
For clarity, we restrict our discussion of this connection to the case treated already in
Example~\ref{ex:Ha} above.
But we believe that the argument is easily generalizable to most Toda-type models.

Let us f\/irst recall that the boundary condition~\eqref{eq:BCtoda} was obtained in~\cite{HK} by analysing
the matrix equation
\begin{gather}
\label{boundary_Hab}
H(m+1,\lambda)A(m,0,\lambda)=A(m,0,h(\lambda))H(m,\lambda),
\end{gather}
where $A(m,n,\lambda)$ is the ``discrete time'' part of the discrete Lax pair (evaluated at the site $n=0$
of the boundary), $h$ is some function acting on the parameter $\lambda$ and $H(m,\lambda)$ is a~matrix
encoding an extra linear symmetry at the site of the boundary (see equations~(14) and (15) in~\cite{HK}) and
ef\/fectively producing allowed integrable boundary conditions.

The main result here is that our zero curvature representation boils down to~\eqref{boundary_Hab} thanks to
a~remarkable ``fusion'' property of the two matrices $L$ which then become $A$ and the fact that
our matrix~$K$ becomes the matrix $H$.
This goes as follows.
The zero curvature representation based on Fig.~\ref{fig:boundaryHAB1} reads
\begin{gather*}
K(x_{m+1};c)L(x_{m+1},q_{m,1},\sigma(a);c)L(q_{m,1},x_m,a;c)
\\
\qquad
=L(x_{m+1},q_{m,1},\sigma(a);\sigma(c))L(q_{m,1},x_m,a;\sigma(c))K(x_m;c),
\end{gather*}
where
\begin{gather*}
L(y,x,a;b)=\frac{1}{(x-y)}\left(
\begin{matrix}
ay+b(x-y) & -axy
\\
a & -ax+b(x-y)
\end{matrix}
\right),
\end{gather*}
and
\begin{gather*}
K(x;c)=\left(
\begin{matrix}
\mu x & (c-\mu)x^2
\\
(c-\mu) & \mu x
\end{matrix}
\right).
\end{gather*}
Now, using the three-leg form
\begin{gather*}
\psi(q_{m,1},x_{m+1};\sigma(a))-\psi(q_{m,1},x_m;a)=\phi(q_{m,1},q_{m,0};\sigma(a),a),
\end{gather*}
and performing calculations similar to those in~\cite[Proposition~11]{BS2}, one f\/inds
\begin{gather*}
L(x_{m+1},q_{m,1},\sigma(a);c)L(q_{m,1},x_m,a;c)=cL(q_{m,0},q_{m,1},\sigma(a)-a;c-a).
\end{gather*}
It remains to perform the changes of parameters as in~\eqref{id}, together with
\begin{gather}
\label{id2}
c\rightarrow(2\lambda+1)b+c\sqrt{\frac{b}{a}},
\end{gather}
to obtain
\begin{gather*}
L(x_{m+1},q_{m,1},\sigma(a);c)L(q_{m,1},x_m,a;c)=2b\left((2\lambda+1)b+c\sqrt{\frac{b}{a}}
\right)\sigma_3A(m,n,\lambda)\sigma_3,
\end{gather*}
where $\sigma_3$ is the usual $2\times 2$ Pauli matrix.
This also yields
\begin{gather*}
K(x_m;c)=\left(
\begin{matrix}
(-1)^m c & (2\lambda+1)a
\\
(2\lambda+1)b & (-1)^m c
\end{matrix}
\right)=(-1)^mH(m,\lambda).
\end{gather*}
This is completely equivalent to the results obtained in~\cite{HK} up to an irrelevant $c\to 1/c$
substitution in $H$.
Note that the map $\lambda\mapsto h(\lambda)=-\lambda-1$ is also correctly reproduced with our choice
$\sigma(c)=-c+2\mu$ and under the identif\/ications~\eqref{id} and~\eqref{id2}.

\section[Connection between 3D boundary consistency and the set-theoretical ref\/lection
equation]{Connection between 3D boundary consistency\\ and the set-theoretical ref\/lection
equation}
\label{sec:conn}

\subsection{General approach}

\begin{figure}[t]
  \centering
\includegraphics{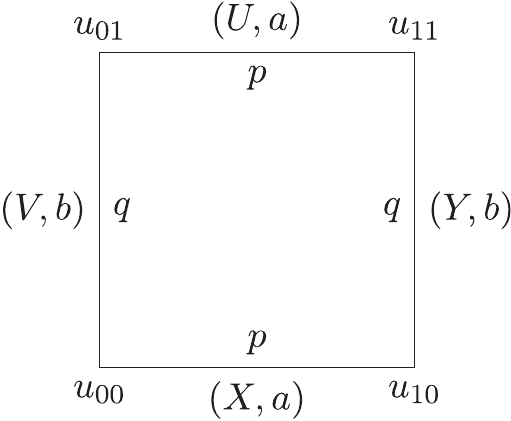}
  \caption{Link between quad-graph equation and Yang--Baxter map.} \label{fig:83}
\end{figure}

In~\cite{PTV}, a~nice approach was described to obtain a~relation between a~3D consistent quad-graph
equation and a~Yang--Baxter map thus yielding a~connection between 3D consistency and the set-theoretical
Yang--Baxter equation~\cite{Drin}.
It is based on the use of symmetries of the equation $Q=0$ and the identif\/ication of invariants under
these symmetries.
Our idea is to extend this connection at the level of ref\/lection maps and integrable boundary equations
thus yielding a~connection between the 3D boundary consistency introduced here and the set-theoretical
ref\/lection equation introduced in~\cite{CCZ, VZ2}.
First, let us recall the method of~\cite{PTV}.
Given a~quad-graph equation
\begin{gather}
\label{eq:8qdef1}
Q(u_{00},u_{10},u_{01},u_{11};a,b)=0,
\end{gather}
let $G$ be a~connected one-parameter group of transformations acting on the variables $u_{ij}$
\begin{gather*}
G_\epsilon: \ (u_{00},u_{10},u_{01},u_{11})\mapsto(\hat{u}_{00},\hat{u}_{10},\hat{u}_{01},\hat{u}_{11}).
\end{gather*}
This transformation is said to be a~symmetry of~\eqref{eq:8qdef1}, if
\begin{gather*}
Q(\hat{u}_{00},\hat{u}_{10},\hat{u}_{01},\hat{u}_{11};a,b)=0,
\end{gather*}
whenever~\eqref{eq:8qdef1} holds.
The corresponding inf\/initesimal action reads
\begin{gather*}
\mathbf{v}Q=0,
\end{gather*}
where
\begin{gather*}
\mathbf{v}=\eta_{00}\frac{\partial}{\partial u_{00}}+\eta_{10}\frac{\partial}{\partial u_{10}}+\eta_{01}
\frac{\partial}{\partial u_{01}}+\eta_{11}\frac{\partial}{\partial u_{11}},
\end{gather*}
with $\eta$ being the characteristic of $G$ specif\/ied by
\begin{gather*}
\eta_{ij}=\frac{d}{d\epsilon}G_\epsilon(u_{ij}).
\end{gather*}
Methods to obtain the characteristic $\eta$ can be seen, e.g.\ in~\cite{levi2007lattice,levi2006continuous}.
Knowing $\mathbf v$, the idea is then to def\/ine a~{\it lattice invariant $I$} of the transformation group
$G$ which satisf\/ies
\begin{gather*}
\mathbf vI=0,
\end{gather*}
and to use it to def\/ine the Yang--Baxter (or edge) variables $(X, Y, U, V)$ from the vertex variables
$(u_{00},u_{10},u_{01},u_{11})$ by
\begin{gather*}
X=I(u_{00},u_{10}),
\qquad
Y=I(u_{10},u_{11}),
\qquad
U=I(u_{01},u_{11}),
\qquad
V=I(u_{00},u_{01}),
\end{gather*}
and assign them to the edges of an elementary quadrilateral as shown in Fig.~\ref{fig:83}.
Once the inf\/initesimal generator $\mathbf v$ is known, one can solve for $I$.

An important result of~\cite{PTV} is that the variables $X$, $Y$, $U$, $V$ are related by a~map
$(U,V)=R(X,Y)$ which is a~Yang--Baxter map, provided the quad-graph equation determined by $Q$ satisf\/ies
the 3D-consistency property.
To construct boundary equations that satisfy the 3D-boundary consistency property, we propose to use this
method ``backwards'' in connection with our classif\/ication of ref\/lection maps associated to
quadrirational Yang--Baxter maps.
Choosing the invariant properly, the Yang--Baxter map~$R$ can be recognized as one canonical form belonging
to the classif\/ication of quadrirational Yang--Baxter maps~\cite{ABS2,PSTV}.
Then we can use the corresponding ref\/lection maps~$h_a$ and $\sigma$ to construct~$q$ according to the
following prescription
\begin{gather}
\label{prescription_q}
q(u_{10},u_{00},u_{01};a)=I(u_{00},u_{01})-h_a(I(u_{00},u_{10})).
\end{gather}
The origin of this prescription comes from the folding method explained and used in~\cite{CCZ} to construct
ref\/lection maps.
When translated in terms of vertex variables $u_{00}$, $u_{10}$, $u_{01}$, it gives~\eqref{prescription_q}.
Indeed, the folding method produces ref\/lection maps $B$ acting on the Yang--Baxter variables and the
parameters, of the form $(V,b)=B(X,a)=(h_a(X),\sigma(a))$.
Hence, $V=h_a(X)$ and recalling that $V = I(u_{00},u_{01})$ and $X = I(u_{00},u_{10})$, this becomes
equivalent to $q(u_{10},u_{00},u_{01};a)=0$ with~$q$ as def\/ined in~\eqref{prescription_q}.
In particular, our construction ensures that $q$ and $Q$ satisfy the 3D-boundary consistency property
since the corresponding Yang--Baxter and ref\/lection maps satisfy the set-theoretical ref\/lection
equation.
We see that to carry out this programme, the knowledge of the invariants and hence~$\mathbf v$ is crucial.
We use the classif\/ication for~$\mathbf v$ obtained in~\cite{rasin2007symmetries} of the so-called
f\/ive-point symmetries, a~subset of which is the set of one-point symmetries which are the ones involved
in the above method.
Note that although the results of~\cite{rasin2007symmetries} were obtained in the context of ${\mathbb
Z}^2$ lattices, we can easily adapt them to the present more general context of quad-graphs.
For each $\mathbf v$ of each quad-graph equation, we provide a~solution~$I_0$ for the invariant~$I$ which
is of the simplest form, the latter meaning that other solutions are obtained from our~$I_0$ in the form
$f(I_0)$ with $f$ being a~dif\/ferentiable bijection depending possibly on the parameters~$a$ or~$b$.
To the best of our knowledge, only sparse examples of invariants and corresponding YB maps have been given
in the literature so far.
In Table~\ref{ta:81} below, we give a~list of invariants satisfying the above criteria and the
corresponding family of YB maps.
Then, one only has to use formula~\eqref{prescription_q} to f\/ind $q$.
In Tables~\ref{ta:Q81} and~\ref{ta:H81}, the solutions for $q$ (and $\sigma$) that we have found using this
method (on top of the folding method) are shown with an asterisk.

\subsection[Example for ${\rm A1}_{\delta=0}$]{Example for $\boldsymbol{{\rm A1}_{\delta=0}}$}

We now carry out the example of the quad-graph equation $A1_{\delta=0}$ explicitly to show how the method
works and illustrate some of the technical subtleties.
Let
\begin{gather*}
Q(u_{00},u_{10},u_{01},u_{11},a,b)=a(u_{00}+u_{01})(u_{10}+u_{11})-b(u_{00}+u_{10})(u_{01}+u_{11}).
\end{gather*}
For this family, the classif\/ication in~\cite{rasin2007symmetries} gives three one-point symmetry
generators\footnote{We have kept the notations of~\cite{rasin2007symmetries} which involve the two integers
$k$, $l$ related to the underlying ${\mathbb Z}^2$ lattice considered in that paper.
However, here, this should be understood as a~straightforward generalization involving the black and white
sublattices to keep track of the relative signs.}
\begin{gather*}
\eta_1=(-1)^{k+l},
\qquad
\eta_2=u_{00},
\qquad
\eta_3=(-1)^{k+l}u^2_{00}.
\end{gather*}
The corresponding simplest invariants $I_1$, $I_2$ and $I_3$ read
\begin{gather*}
I_1(s,t)=s+t,
\qquad
I_2(s,t)=\frac{s}{t},
\qquad
I_3(s,t)=\frac{1}{s}+\frac{1}{t}.
\end{gather*}
So, for the f\/irst invariant, the Yang--Baxter variables read
\begin{gather*}
X=u_{00}+u_{10},
\qquad
Y=u_{10}+u_{11},
\qquad
U=u_{01}+u_{11},
\qquad
V=u_{00}+u_{01},
\end{gather*}
which satisfy
\begin{gather}
\label{relations_F3}
X-Y=V-U,
\qquad
aVY-bXU=0.
\end{gather}
This yields the relations
\begin{gather*}
U=aY\frac{X-Y}{bX-aY},
\qquad
V=bX\frac{X-Y}{bX-aY}.
\end{gather*}
To recognize the family to which this map belongs, we perform the following transformation on the variables
\begin{gather}
\label{transfo_1}
X\to a X,
\qquad
Y\to b Y,
\qquad
U\to a U,
\qquad
V\to b V.
\end{gather}
We then obtain the $F_{\rm III}$ quadrirational Yang--Baxter map
\begin{gather*}
U=\frac{Y}{a}P,
\qquad
V=\frac{X}{b}P,
\qquad
P=\frac{aX-bY}{X-Y}.
\end{gather*}
For this family, we have the following ref\/lections maps
\begin{gather*}
\sigma(a)=\frac{\mu^2}{a}\qquad \text{and}\qquad h_a(X)=\frac{a X}{\mu}\qquad \text{or}\qquad h_a(X)=-\frac{a X}{\mu},
\end{gather*}
where $\mu$ is a~free parameter.
Performing the inverse transformation of~\eqref{transfo_1}, we obtain the ref\/lection maps that we can use
in formula~\eqref{prescription_q} to obtain $q$.
They read
\begin{gather*}
h_a(X)=\frac{\mu X}{a}\qquad \text{or}\qquad h_a(X)=-\frac{\mu X}{a},
\end{gather*}
and we obtain
\begin{gather*}
q(x,y,z;a)=\mu(x+y)-a(y+z)\qquad \text{or}\qquad q(x,y,z;a)=\mu(x+y)+a(y+z),
\end{gather*}
both valid with $\sigma(a)=\frac{\mu^2}{a}$.
Note that in this case, the two possibilities for the boundary equations $q(u_{10},u_{00},u_{01};a)=0$ are
related by the transformation $\mu\to -\mu$ which leaves $\sigma(a)$ invariant so that, in fact, we only
have one boundary equation here.

Let us perform the same analysis for $I_2$.
The Yang--Baxter variables are
\begin{gather*}
X=u_{00}/u_{10},
\qquad
Y=u_{10}/u_{11},
\qquad
U=u_{01}/u_{11},
\qquad
V=u_{00}/u_{01},
\end{gather*}
and they satisfy
\begin{gather*}
XY=UV,
\qquad
a(XY+U)(Y+1)-bY(X+1)(U+1)=0,
\end{gather*}
which yields
\begin{gather*}
U=\frac{b Y-a X Y+b X Y-a X Y^2}{a+a Y-b Y-b X Y},
\qquad
V=\frac{a X+a X Y-b X Y-b X^2Y}{b-a X+b X-a X Y}.
\end{gather*}
Performing the following transformation
\begin{gather}
\label{transfo_2}
X\to-X,
\qquad
Y\to-Y,
\qquad
U\to-U,
\qquad
V\to-V,
\end{gather}
we obtain the $H_{\rm II}$ quadrirational Yang--Baxter map
\begin{gather*}
U=Y\frac{(b+(a-b)X-a XY)}{(a+(b-a)Y-b XY)},
\qquad
V=X\frac{(a+(b-a)Y-b XY)}{(b+(a-b)X-a XY)}.
\end{gather*}
For this family, we have the following ref\/lections maps
\begin{gather*}
\sigma(a)=\frac{\mu^2}{a}\qquad \text{and}\qquad h_a(X)=\frac{a+\mu-X\mu}{a}\qquad \text{or}\qquad h_a(X)=\frac{a X}
{a X+\mu-X\mu},
\end{gather*}
or,
\begin{gather*}
\sigma(a)=-a+2\mu\qquad \text{and}\qquad h_a(X)=-X\qquad \text{or}\qquad h_a(X)=\frac{a+(X-1)\mu}{a X+\mu-X\mu}.
\end{gather*}
where $\mu$ is a~free parameter.
Performing the inverse transformation of~\eqref{transfo_2} and using~\eqref{prescription_q} we f\/ind
\begin{gather*}
q(x,y,z;a)=a x(y+z)+(x+y)z\mu\qquad \text{or}\qquad q(x,y,z;a)=y(a(y+z)-(x+y)\mu),
\end{gather*}
both valid with $\sigma(a)=\frac{\mu^2}{a}$, and,
\begin{gather*}
q(x,y,z;a)=y(x+z)\qquad \text{or}\qquad q(x,y,z;a)=a\left(y^2-x z\right)-(x+y)(y-z)\mu,
\end{gather*}
both valid for $\sigma(a)=-a+2 \mu$.

Finally, using $I_3$ yields the relations~\eqref{relations_F3} so that we obtain the $F_{\rm III}$
quadrirational Yang--Baxter map again.
Therefore, we can use the same ref\/lection maps but because the invariant is dif\/ferent we may obtain
dif\/ferent expressions for $q$.
A direct calculation gives
\begin{gather*}
q(x,y,z;a)=a x(y+z)+(x+y)z\mu
\qquad
\text{or}
\qquad
q(x,y,z;a)=a x(y+z)-(x+y)z\mu,
\end{gather*}
both valid with $\sigma(a)=\frac{\mu^2}{a}$.
These are in fact the same solution under the transformation $\mu\to -\mu$ and it coincides with the
f\/irst solution already obtained from the $H_{\rm II}$ family.
Let us remark that in some cases, point transformations on the parameters are needed on top of M\"obius
transformations to recognize the canonical Yang--Baxter map.
This may af\/fect the form of the map $\sigma$ to be used in the 3D boundary consistency condition.
This happens for ${\rm Q3}_{\delta=0}$, $\rm H3$ and $\rm A2$.
Finally, let us also mention that this method has some inherent limitations due to the fact that for
families ${\rm Q2}$, ${\rm Q3}_{\delta=1}$ and $\rm Q4$, there are no one-point symmetry generators in the classif\/ication
of~\cite{rasin2007symmetries}.
Our other method described in Section~\ref{sec:class} does also work for these families, as can be
seen in the tables.
Whether or not the corresponding solutions for $q$ for these ``missing'' families can be mapped back to
ref\/lection maps is an interesting open question.
\begin{table}[t!]\centering
\caption{Some invariants and corresponding Yang--Baxter maps for the ABS quad-graph equations.\label{ta:81}}
\vspace{2mm}

\begin{tabular}{@{}l@{\,\,\,\,\,}l@{\,\,\,\,\,}l@{\,\,\,\,\,}l@{}}
\hline
Quad-graph equation & Characteristic &Invariant & Yang--Baxter map\tsep{2pt}\bsep{2pt}
\\
\hline
\multirow{3}{*}{Q1$_{\delta=0}$ } & $\eta _1 = 1$ & $I(s,t)=s-t$ & $
H_{{\rm III}a}$\tsep{2pt}\bsep{2pt}
\\
& $\eta _2 = u_{00} $ & $I(s,t)=s/t$ & $H_{\rm II}$\bsep{2pt}
\\
& $\eta _3 = u^2_{00} $ & $I(s,t)=1/s-1/t$ & $H_{{\rm III}a}$\bsep{2pt}
\\
\hline
Q1$_{\delta=1}$ & $\eta_1 = 1$ & $I(s,t)=s-t$ & $H_{\rm II}$\tsep{2pt}\bsep{2pt}
\\
\hline
Q3$_{\delta=0}$ & $\eta _1 = u_{00}$ & $I(s,t)=s/t$ & $H_{\rm I}$\tsep{2pt}\bsep{2pt}
\\
\hline
\multirow{3}{*}{H1} & $\eta _1 = 1$ & $I(s,t)=s-t$ & $H_{\rm V}$\tsep{2pt}\bsep{2pt}
\\
& $\eta _2 = (-1)^{k+l}$ & $I(s,t)=s+t$ & $F_{\rm V}$\bsep{2pt}
\\
& $\eta _3 = (-1)^{k+l}u_{00}$ & $I(s,t)=st$ & $F_{\rm IV}$\bsep{2pt}
\\
\hline
H2 & $\eta _1 = (-1)^{k+l}$ & $I(s,t)=s+t$ & $F_{\rm IV}$\tsep{2pt}\bsep{2pt}
\\
\hline
\multirow{2}{*}{H3$_{\delta=0}$} & $\eta _1 = u_{00}$ & $I(s,t)=s/t$ & $H_{{\rm III}b}$\tsep{2pt}\bsep{2pt}
\\
& $\eta _2 = (-1)^{k+l} u_{00} $ & $I(s,t)=st$ & $F_{\rm III}$\bsep{2pt}
\\
\hline
H3$_{\delta=1}$ & $\eta _1 = (-1)^{k+l} u_{00}$ & $I(s,t)=st$ & $F_{\rm II}$\tsep{2pt}\bsep{2pt}
\\
\hline
\multirow{3}{*}{A1$_{\delta=0}$ } & $\eta _1 = (-1)^{k+l}$ & $I(s,t)=s+t$ & $F_{\rm III}$\tsep{2pt}\bsep{2pt}
\\
& $\eta _2 = u_{00} $ & $I(s,t)=s/t$ & $H_{\rm II}$\bsep{2pt}
\\
& $\eta _3 = (-1)^{k+l} u^2_{00} $ & $I(s,t)=1/s+1/t$ & $F_{\rm III}$\bsep{2pt}
\\
\hline
A1$_{\delta=1}$ & $\eta _1 = (-1)^{k+l} $ & $I(s,t)=s+t$ & $F_{\rm II}$\tsep{2pt}\bsep{2pt}
\\
\hline
A2 & $\eta _1 = (-1)^{k+l} u_{00}$ & $I(s,t)=st$ & $F_{\rm I}$\tsep{2pt}\bsep{2pt}
\\
\hline
\end{tabular}
\end{table}

\section{Conclusions and outlooks}

We proposed a~def\/inition for integrable equations on a~quad-graph with
boundary and introduced the notion of 3D boundary consistency as
a~complement of the 3D consistency condition that is used as an integrability criterion
for bulk quad-graph systems.
Just like quadrilaterals are fundamental structures when one discretizes an arbitrary surface without
boundary, we argued that triangles appear naturally when one considers surfaces with boundary.
Therefore, it is natural to associate a~$3$-point boundary equation to describe the boundary locally just
like one associates a~$4$-point bulk equation to describe the bulk locally.
We presented two dif\/ferent methods to f\/ind solutions of the 3D boundary consistency condition given
an integrable quad-graph equation of the ABS classif\/ication.
The terminology ``integrable boundary'' is also backed up by the discussion of other traditional integrable
structures like B\"acklund transformations and zero curvature representation for systems with boundary.
This is also supported by the connection that we unraveled between 3D boundary consistency and
set-theoretical ref\/lection equation.
As a~by-product of our study, we were also able to introduce three-leg forms of boundary equations and
hence to introduce Toda-type systems with boundary.

The present work lays some foundations for what we hope could be a~new exciting area of research in
discrete integrable systems.
Among the many open questions one can think of, we would like to mention a~few that we believe are
important: f\/inding a~method of classif\/ication of boundary equations given a~bulk quad-graph equation,
tackling the problem of posing the initial-boundary value problem for quad-graph equations with a~boundary,
understanding the connection of our approach with the singular-boundary reduction approach of~\cite{AJ},
implementing the discrete inverse scattering method with a~boundary with the hope of f\/inding soliton
solutions, etc.

\subsection*{Acknowledgements}

The f\/inal details of this paper were completed while two of the authors (V.C.\ and Q.C.Z) were at the
``Discrete Integrable Systems'' conference held at the Newton Institute for Ma\-the\-ma\-ti\-cal Sciences.
We wish to thank C.~Viallet for pointing out useful references.
We also thank M.~Nieszporski and P.~Kassotakis for useful discussions and the provision of unpublished
material on their work on the connection between Yang--Baxter maps and quad-graph equations, some of which
is related to our results shown in Table~\ref{ta:81}.
Last, but not least, we express our sincere gratitude to the referees whose excellent comments and
criticisms helped improve this paper tremendously.

\pdfbookmark[1]{References}{ref}
\LastPageEnding


\begin{thebibliography}{99}
\footnotesize\itemsep=0pt

\bibitem{Adl}
Adler V.E., Discrete equations on planar graphs, \href{http://dx.doi.org/10.1088/0305-4470/34/48/310}{\textit{J.~Phys.~A: Math.
  Gen.}} \textbf{34} (2001), 10453--10460.

\bibitem{ABS}
Adler V.E., Bobenko A.I., Suris Yu.B., Classif\/ication of integrable equations on
  quad-graphs. {T}he consistency approach, \textit{Comm. Math. Phys.}
  \textbf{233} (2003), 513--543, \href{http://arxiv.org/abs/nlin.SI/0202024}{nlin.SI/0202024}.

\bibitem{ABS2}
Adler V.E., Bobenko A.I., Suris Yu.B., Geometry of {Y}ang--{B}axter maps:
  pencils of conics and quadrirational mappings, \textit{Comm. Anal. Geom.}
  \textbf{12} (2004), 967--1007, \href{http://arxiv.org/abs/math.QA/0307009}{math.QA/0307009}.

\bibitem{AK}
Ahn C., Koo W.M., Boundary {Y}ang--{B}axter in the {RSOS}/{SOS} representation,
  in Statistical Models, {Y}ang--{B}axter Equation and Related Topics, and
  {S}ymmetry, Statistical Mechanical Models and Applications ({T}ianjin, 1995),
  World Sci. Publ., River Edge, NJ, 1996, 3--12, \href{http://arxiv.org/abs/hep-th/9508080}{hep-th/9508080}.

\bibitem{AHN}
Atkinson J., Hietarinta J., Nijhof\/f F., Seed and soliton solutions for
  {A}dler's lattice equation, \href{http://dx.doi.org/10.1088/1751-8113/40/1/F01}{\textit{J.~Phys.~A: Math. Theor.}} \textbf{40}
  (2007), F1--F8, \href{http://arxiv.org/abs/nlin.SI/0609044}{nlin.SI/0609044}.

\bibitem{AJ}
Atkinson J., Joshi N., Singular-boundary reductions of type-{Q} {ABS}
  equations, \href{http://dx.doi.org/10.1093/imrn/rns024}{\textit{Int. Math. Res. Not.}} \textbf{2013} (2013), 1451--1481,
  \href{http://arxiv.org/abs/1108.4502}{arXiv:1108.4502}.

\bibitem{Bax86}
Baxter R.J., The {Y}ang--{B}axter equations and the {Z}amolodchikov model,
  \href{http://dx.doi.org/10.1016/0167-2789(86)90195-8}{\textit{Phys.~D}} \textbf{18} (1986), 321--347.

\bibitem{BMS}
Bazhanov V.V., Mangazeev V.V., Sergeev S.M., Quantum geometry of
  three-dimensional lattices, \href{http://dx.doi.org/10.1088/1742-5468/2008/07/P07004}{\textit{J.~Stat. Mech. Theory Exp.}} \textbf{2008}
  (2008), P07004, 27~pages, \href{http://arxiv.org/abs/0801.0129}{arXiv:0801.0129}.

\bibitem{BPO}
Behrend R.E., Pearce P.A., O'Brien D.L., Interaction-round-a-face models with
  f\/ixed boundary conditions: the {ABF} fusion hierarchy, \href{http://dx.doi.org/10.1007/BF02179576}{\textit{J.~Statist.
  Phys.}} \textbf{84} (1996), 1--48, \href{http://arxiv.org/abs/hep-th/9507118}{hep-th/9507118}.

\bibitem{VB}
Bellon M.P., Viallet C.-M., Algebraic entropy, \href{http://dx.doi.org/10.1007/s002200050652}{\textit{Comm. Math. Phys.}}
  \textbf{204} (1999), 425--437, \href{http://arxiv.org/abs/chao-dyn/9805006}{chao-dyn/9805006}.

\bibitem{BS2}
Bobenko A.I., Suris Yu.B., Integrable systems on quad-graphs, \href{http://dx.doi.org/10.1155/S1073792802110075}{\textit{Int. Math.
  Res. Not.}} \textbf{2002} (2002), 573--611, \href{http://arxiv.org/abs/nlin.SI/0110004}{nlin.SI/0110004}.

\bibitem{BS}
Bobenko A.I., Suris Yu.B., Discrete dif\/ferential geometry. Integrable structure,
  \href{http://dx.doi.org/10.1007/978-3-7643-8621-4}{\textit{Graduate Studies in Mathematics}}, Vol.~98, American Mathematical
  Society, Providence, RI, 2008, \href{http://arxiv.org/abs/math.DG/0504358}{math.DG/0504358}.

\bibitem{Boll}
Boll R., Classif\/ication of 3{D} consistent quad-equations, \href{http://dx.doi.org/10.1142/S1402925111001647}{\textit{J.~Nonlinear
  Math. Phys.}} \textbf{18} (2011), 337--365, \href{http://arxiv.org/abs/1009.4007}{arXiv:1009.4007}.

\bibitem{CCZ}
Caudrelier V., Cramp{\'e} N., Zhang Q.C., Set-theoretical ref\/lection equation:
  classif\/ication of ref\/lection maps, \href{http://dx.doi.org/10.1088/1751-8113/46/9/095203}{\textit{J.~Phys.~A: Math. Theor.}}
  \textbf{46} (2013), 095203, 12~pages, \href{http://arxiv.org/abs/1210.5107}{arXiv:1210.5107}.

\bibitem{VZ2}
Caudrelier V., Zhang Q.C., Yang--Baxter and ref\/lection maps from vector
  solitons with a boundary, \href{http://arxiv.org/abs/1205.1133}{arXiv:1205.1133}.

\bibitem{Che}
Cherednik I.V., Factorizing particles on a half-line and root systems,
  \href{http://dx.doi.org/10.1007/BF01038545}{\textit{Theoret. and Math. Phys.}} \textbf{61} (1984), 977--983.

\bibitem{Drin}
Drinfeld V.G., On some unsolved problems in quantum group theory, in Quantum
  Groups ({L}eningrad, 1990), \href{http://dx.doi.org/10.1007/BFb0101175}{\textit{Lecture Notes in Math.}}, Vol.~1510,
  Springer, Berlin, 1992, 1--8.

\bibitem{FHS}
Fan H., Hou B.Y., Shi K.J., General solution of ref\/lection equation for
  eight-vertex {SOS} model, \href{http://dx.doi.org/10.1088/0305-4470/28/17/010}{\textit{J.~Phys.~A: Math. Gen.}} \textbf{28} (1995),
  4743--4749.

\bibitem{GRP}
Grammaticos B., Ramani A., Papageorgiou V., Do integrable mappings have the
  {P}ainlev\'e property?, \href{http://dx.doi.org/10.1103/PhysRevLett.67.1825}{\textit{Phys. Rev. Lett.}} \textbf{67} (1991),
  1825--1828.

\bibitem{HK}
Habibullin I.T., Kazakova T.G., Boundary conditions for integrable discrete
  chains, \href{http://dx.doi.org/10.1088/0305-4470/34/48/303}{\textit{J.~Phys.~A: Math. Gen.}} \textbf{34} (2001), 10369--10376.

\bibitem{Hie}
Hietarinta J., Searching for {CAC}-maps, \href{http://dx.doi.org/10.2991/jnmp.2005.12.s2.16}{\textit{J.~Nonlinear Math. Phys.}}
  \textbf{12} (2005), suppl.~2, 223--230.

\bibitem{levi2007lattice}
Levi D., Petrera M., Scimiterna C., The lattice {S}chwarzian {K}d{V} equation
  and its symmetries, \href{http://dx.doi.org/10.1088/1751-8113/40/42/S18}{\textit{J.~Phys.~A: Math. Theor.}} \textbf{40} (2007),
  12753--12761, \href{http://arxiv.org/abs/math-ph/0701044}{math-ph/0701044}.

\bibitem{levi2006continuous}
Levi D., Winternitz P., Continuous symmetries of dif\/ference equations,
  \href{http://dx.doi.org/10.1088/0305-4470/39/2/R01}{\textit{J.~Phys.~A: Math. Gen.}} \textbf{39} (2006), R1--R63,
  \href{http://arxiv.org/abs/nlin.SI/0502004}{nlin.SI/0502004}.

\bibitem{mercatthesis}
Mercat C., Holomorphie discr\`ete et mod\`ele d'Ising, Ph.D. Thesis,
  Universit\'e Louis Pasteur, Strasbourg, France, 1998, available at
  \url{http://tel.archives-ouvertes.fr/tel-00001851/}.

\bibitem{M}
Mercat C., Discrete {R}iemann surfaces and the {I}sing model, \href{http://dx.doi.org/10.1007/s002200000348}{\textit{Comm.
  Math. Phys.}} \textbf{218} (2001), 177--216.

\bibitem{Nij}
Nijhof\/f F.W., Lax pair for the {A}dler (lattice {K}richever--{N}ovikov) system,
  \href{http://dx.doi.org/10.1016/S0375-9601(02)00287-6}{\textit{Phys. Lett.~A}} \textbf{297} (2002), 49--58, \href{http://arxiv.org/abs/nlin.SI/0110027}{nlin.SI/0110027}.

\bibitem{PSTV}
Papageorgiou V.G., Suris Yu.B., Tongas A.G., Veselov A.P., On quadrirational
  {Y}ang--{B}axter maps, \href{http://dx.doi.org/10.3842/SIGMA.2010.033}{\textit{SIGMA}} \textbf{6} (2010), 033, 9~pages,
  \href{http://arxiv.org/abs/0911.2895}{arXiv:0911.2895}.

\bibitem{PTV}
Papageorgiou V.G., Tongas A.G., Veselov A.P., Yang--{B}axter maps and
  symmetries of integrable equations on quad-graphs, \href{http://dx.doi.org/10.1063/1.2227641}{\textit{J.~Math. Phys.}}
  \textbf{47} (2006), 083502, 16~pages, \href{http://arxiv.org/abs/math.QA/0605206}{math.QA/0605206}.

\bibitem{rasin2007symmetries}
Rasin O.G., Hydon P.E., Symmetries of integrable dif\/ference equations on the
  quad-graph, \href{http://dx.doi.org/10.1111/j.1467-9590.2007.00385.x}{\textit{Stud. Appl. Math.}} \textbf{119} (2007), 253--269.

\bibitem{Sk}
Sklyanin E.K., Boundary conditions for integrable quantum systems,
  \href{http://dx.doi.org/10.1088/0305-4470/21/10/015}{\textit{J.~Phys.~A: Math. Gen.}} \textbf{21} (1988), 2375--2389.

\end{thebibliography}
\end{document}